\documentclass[a4paper,fleqn,usenatbib]{mnras}

\usepackage{newtxtext,newtxmath}
\usepackage[T1]{fontenc}

\usepackage[english]{babel}
\usepackage{graphicx}
\usepackage{fleqn}
\usepackage{amssymb}
\usepackage{amsmath}
\usepackage{booktabs}
\usepackage{hyperref}
\usepackage{comment}
\usepackage{enumitem}

\setlist[enumerate]{
  labelsep=8pt,
  labelindent=0.\parindent,
 itemindent=0pt,
  leftmargin=*,
}

\usepackage{color, colortbl}

\newcommand*{\mycdot}{\kern-.2em\cdot\kern-.2em}
\renewcommand{\S}{Section}
\newcommand{\Ss}{Sections}
\newcommand{\F}{Fig.}
\newcommand{\eq}{equation}
\newcommand{\eqs}{equations}

\newcommand{\ve}[1]{\boldsymbol{#1}}
\newcommand{\unit}[1]{\hat{\boldsymbol{#1}}}

\newcommand{\au}{\,\textsc{au}}
\newcommand{\yr}{\mathrm{yr}}
\newcommand{\kyr}{\mathrm{kyr}}
\newcommand{\Myr}{\mathrm{Myr}}

\newcommand{\sm}{\textsc{SecularMultiple}}

\renewcommand{\j}{\jmath}

\newcommand{\Mpn}{\mathcal{M}_p^{(n)}}
\newcommand{\Anm}{\mathcal{A}_m^{(n)}}
\newcommand{\Bnmi}{\mathcal{B}_{i_1,i_2}^{(n,m)} (e_p)}
\newcommand{\Bnmie}{\mathcal{B}_{i_1,i_2}^{(n,m)} (e)}
\newcommand{\Cnil}{\mathcal{C}^{(n,i_1,i_2)}_{l_1,l_2,l_3,l_4}(e_p,e_k)}

\newcommand{\vdot}[2]{\left ( #1 \cdot #2 \right )}

\pubyear{2020}

\usepackage{listings}

\usepackage{color}

\definecolor{dkgreen}{rgb}{0,0.6,0}
\definecolor{gray}{rgb}{0.5,0.5,0.5}
\definecolor{mauve}{rgb}{0.58,0,0.82}

\allowdisplaybreaks

\begin{document}
\title[Suborbital effects in hierarchical systems]{Secular dynamics of hierarchical multiple systems composed of nested binaries, with an arbitrary number of bodies and arbitrary hierarchical structure. III. Suborbital effects: hybrid integration techniques and orbit-averaging corrections}

\author[Hamers]{Adrian S. Hamers$^{1}$\thanks{E-mail: hamers@mpa-garching.mpg.de}\\
$^{1}$Max-Planck-Institut f\"{u}r Astrophysik, Karl-Schwarzschild-Str. 1, 85741 Garching, Germany}
\date{Accepted 2020 April 16. Received 2020 April 16; in original form 2020 March 13.}

\label{firstpage}
\pagerange{\pageref{firstpage}--\pageref{lastpage}}
\maketitle

\begin{abstract}  
The \sm~code, presented in two previous papers of this series, integrates the long-term dynamical evolution of multiple systems with any number of bodies and hierarchical structure, provided that the system is composed of nested binaries. In the formalism underlying \sm, we previously averaged over all orbits in the system. This approximation significantly speeds up numerical integration of the equations of motion, making large population synthesis studies possible. However, the orbit averaging approximation can break down when the secular evolution timescale of the system is comparable to or shorter than any of the orbital periods in the system. Here, we present an update to \sm~in which we incorporate hybrid integration techniques, and orbit-averaging corrections. With this update, the user can specify which orbits should be integrated directly (without averaging), or assuming averaged orbits. For orbits that are integrated directly, we implemented two integration techniques, one which is based on the regularised Kustaanheimo-Stiefel equations of motion in element form. We also implemented analytical orbit-averaging corrections for pairwise interactions to quadrupole order. The updates presented here provide more flexibility for integrating the long-term dynamical evolution of hierarchical multiple systems. By effectively combining direct integration and orbit averaging the long-term evolution can be accurately computed, but with significantly lower computational cost compared to existing direct $N$-body codes. We give a number of examples in which the new features are beneficial. Our updated code, which is written in \textsc{C++} supplemented with a user-friendly interface in \textsc{Python}, is freely available.
\end{abstract}


\section{Introduction}
\label{sect:introduction}
Owing to their long-term dynamical stability, hierarchies appear in a plethora in astrophysical systems. The Solar system can be regarded as a hierarchical system (with each planet orbiting the centre of mass of the subsystem inside it), and the thousands of exoplanet systems that have been discovered to date show a rich variety of orbital architectures (see, e.g., \citealt{2015ARA&A..53..409W}), including planets in multiple-star systems such as binaries, triples, and even quadruples (e.g., PH1, \citealt{2013ApJ...768..127S}, and 30 Arietis, \citealt{2009A&A...507.1659G}). Stellar systems such as triple- and quadruple-star systems are hierarchical systems themselves, and are common. Among Solar-type stars, triples and quadruples comprise approximately 10\% and 1\% of stellar systems, respectively \citep{2010ApJS..190....1R,2014AJ....147...86T,2014AJ....147...87T}. For more massive systems, the triple and quadruple fractions are significantly higher, and are each roughly 40\% among systems with O-star primaries \citep{2017ApJS..230...15M}. 

The simplest type of hierarchical system, a hierarchical triple, can display secular oscillations of the inner orbit eccentricity known as Lidov-Kozai (LK) oscillations\footnote{It has recently been noted that Hugo von Zeipel \citep{1910AN....183..345V} made important contributions to this topic well before Lidov and Kozai \citep{2019MEEP....7....1I}.} \citep{1962P&SS....9..719L,1962AJ.....67..591K}. During these oscillations, large inner orbit eccentricities can be attained, which can have important implications in a variety of contexts such as producing short-period binaries (e.g., \citealt{1979A&A....77..145M,1998MNRAS.300..292K,2001ApJ...562.1012E,2006Ap&SS.304...75E,2007ApJ...669.1298F,2014ApJ...793..137N}) and hot Jupiters (e.g., \citealt{2003ApJ...589..605W,2007ApJ...669.1298F,2012ApJ...754L..36N,2015ApJ...799...27P,2016MNRAS.456.3671A,2016ApJ...829..132P}), enhancing mergers of compact objects (e.g., \citealt{2002ApJ...578..775B,2011ApJ...741...82T,2013MNRAS.430.2262H,2017ApJ...841...77A,2017ApJ...836...39S,2017ApJ...846L..11L,2018ApJ...863...68L,2018ApJ...856..140H,2018ApJ...853...93R,2018ApJ...864..134R,2018A&A...610A..22T,2019MNRAS.486.4443F}), affecting the evolution of protoplanetary or accretion disks in binaries (e.g., \citealt{2014ApJ...792L..33M,2015ApJ...813..105F,2017MNRAS.467.1957Z,2017MNRAS.469.4292L,2018MNRAS.477.5207Z,2019MNRAS.485..315F,2019MNRAS.489.1797M}), and producing blue straggler stars (e.g., \citealt{2009ApJ...697.1048P,2016ApJ...816...65A,2016MNRAS.460.3494S,2019MNRAS.488..728F}).

Although the dynamics of hierarchical triple systems can be intricate, they become even more complex with the addition of another body in hierarchical quadruple systems. More specifically, (long-term stable) quadruples, which appear in the 3+1 or 2+2 configurations, can give rise to stronger secular evolution (i.e., leading to higher eccentricities) in a larger parameter space \citep{2013MNRAS.435..943P,2015MNRAS.449.4221H,2016MNRAS.461.3964V,2017MNRAS.470.1657H,2018MNRAS.476.4234F,2018MNRAS.474.3547G,2019MNRAS.483.4060L,2019MNRAS.486.4781F}. Even higher-multiplicity systems (quintuples, sextuples, etc.) can similarly give rise to strong secular evolution \citep{2020arXiv200208746H}.

The long-term dynamical evolution of hierarchical systems can be computed straightforwardly using direct $N$-body integration. However, as a result of the often large separation between orbital timescales and secular evolution timescales, such integrations can be computationally costly. In two previous papers (\citealt{2016MNRAS.459.2827H}, hereafter Paper I; \citealt{2018MNRAS.476.4139H}, hereafter Paper II), we presented a formalism and algorithm (named \sm) to efficiently model the secular evolution of hierarchical systems composed of nested orbits, with arbitrary structure and number of bodies. The formalism is based on an expansion of the Hamiltonian of the system in terms of ratios of the separations of all binaries in the system. Subsequently, the Hamiltonian is averaged over all orbits in the system. This approach is a generalisation of a commonly-used technique in hierarchical triples, where an expansion of the Hamiltonian is made in terms of the small quantity $r_\mathrm{in}/r_\mathrm{out}$, where $r_\mathrm{in}$ and $r_\mathrm{out}$ are the (instantaneous) separations of the inner and outer orbits, and the inner and outer orbits are subsequently averaged over (e.g., \citealt{1962P&SS....9..719L,1962AJ.....67..591K,1968AJ.....73..190H,2000ApJ...535..385F,2013MNRAS.431.2155N}). 

The expansion approximation is valid as long as $x_{ij} = r_i/r_j$, where $i$ refers to any inner orbit and $j$ to any outer orbit in the system, satisfies $x_{ij} \ll 1$. The averaging approximation holds when the secular evolution timescales in the system are (significantly) longer than the any of the orbital periods in the system. Although these approximations are justified in many cases, there are others in which they break down. When some or all of the $x_{ij}$ in the system are not small and approach unity, the system is likely to become dynamically unstable. Such a (typically short-lived) dynamical instability phase, which in practice can be triggered by processes such as stellar evolution (e.g., \citealt{2012ApJ...760...99P}), could lead to ejections of bodies from the system, destroying the hierarchy of the system. 

A less extreme possibility is that the averaging approximation breaks down. In this case, the hierarchy of the system can remain intact, yet the dynamical interactions can be strong enough that the secular evolution timescale is shorter than any of the orbital periods (usually the longest orbital period). This breakdown of the orbit averaging approximation has been studied in recent years for hierarchical triples by various authors \citep{2012ApJ...757...27A,2014ApJ...781...45A,2016MNRAS.458.3060L,2018MNRAS.481.4907G,2018MNRAS.481.4602L,2019MNRAS.490.4756L}, who have shown that perturbations on orbital timescales can accumulate, and affect the long-term evolution of the system. Methods to include these suborbital effects include averaging the inner orbit but not the outer orbit, i.e., integrating the outer orbit directly (e.g., \citealt{2016PhDT........82A}), and analytically deriving `corrections' to the double averaging terms (e.g., \citealt{2016MNRAS.458.3060L,2018MNRAS.481.4602L,2019MNRAS.490.4756L}). 

In this paper, we present an update to the \sm~code in which we take into account suborbital effects with two complementary approaches. First, we implemented hybrid integration techniques, which allow the user of the code to specify which orbits in the system should be averaged over, and which ones should be integrated directly. This approach is motivated by the fact that, especially in high-multiplicity hierarchical systems, there can be situations in which the orbit averaging approximation is well justified for one or more inner orbits, but not all outer orbits. Such situations call for an integration scheme in which the fast inner orbits are averaged over, but not the slower outer orbits. For consistency, in our algorithm we only allow `inner' orbits to be averaged, and `outer' orbits to be integrated directly (for example, in a 3+1 quadruple, the two inner orbits could be averaged over and the outermost orbit integrated directly; averaging over the outer orbit but directly integrating the inner two orbits, although technically possible, is not allowed since it would not be self-consistent). This hybrid integration technique provides flexibility in integrating the long-term dynamical evolution of hierarchical multiple systems. When applied appropriately depending on the system, a significant speedup compared to direct $N$-body integration can be attained whereas still retaining similar accuracy, i.e., the important suborbital effects are still taken into account. 

Second, we implemented analytical orbit-averaging corrections within the \sm~code. Currently, we implemented orbit averaging terms to the quadrupole expansion order and valid at the test particle approximation in triples (in which one of the bodies in the inner binary is massless, such that the outer orbit is static), as derived by \citet{2016MNRAS.458.3060L}. A self-consistent derivation of averaging corrections in higher-multiplicity systems and in the general (non-test-particle) case is left for future work. 

The \sm~code is written in \textsc{C++} and has a user-friendly \textsc{Python} interface. It is freely available online\footnote{\href{https://github.com/hamers/secularmultiple}{https://github.com/hamers/secularmultiple}}, where test and example scripts are also provided. 

The structure of this paper is as follows. In \S~\ref{sect:meth}, we describe the methodology of hybrid integration within \sm, and illustrate its use in practice. Also, we discuss the orbit-averaging corrections implemented in the code. In \S~\ref{sect:use}, we illustrate the use of the new features in \sm~in practice. In \S~\ref{sect:ex}, we present a number of examples in which the added features in \sm~can be beneficial. We discuss our results in \S~\ref{sect:discussion}, and conclude in \S~\ref{sect:conclusions}.

\begin{table}
\begin{tabular}{lp{5.6cm}}
\toprule
Symbol & Description \\
\midrule
$G$ & Gravitational constant. \\
$H$ & Hamiltonian of the system (unavaraged). \\
$H_\mathrm{Kep}$ & Keplerian part of the Hamiltonian (see \eq~\ref{eq:Hkep}). \\
$k \in \mathrm{B}$ & Iteration over all binaries in the system ($\mathrm{B}$ is the set of all binaries). \\
$S'_n$ & Perturbing potential of order $n$ (see \eq~\ref{eq:Spn}). \\
$S'_{n;j}$ & Perturbing potential of order $n$ for interactions involving $j$ binaries. \\
$\langle (...) \rangle_k$ & Quantity (...) averaged over orbit $k$. \\
$\ve{r}_k$ & Relative separation vector of orbit $k$. \\
$M_k$ & Mass of all bodies contained within orbit $k$. \\
$M_{k.\mathrm{C}j}$ &Mass of all bodies contained within child $j$ of orbit $k$ ($j$ can be either 1 or 2).  \\
$a_k$ & Semimajor axis of orbit $k$. \\
$e_k$ & Eccentricity of orbit $k$. \\
$\ve{e}_k$ & Eccentricity vector of orbit $k$. \\
$\ve{\j}_k$ & Dimensionless angular-momentum vector of orbit $k$; its magnitude is $\j_k = \sqrt{1-e_k^2}$. \\
$\mu_k$ & Reduced mass of orbit $k$ (see \eq~\ref{eq:mu}). \\
$\Lambda_k$ & Circular angular momentum of orbit $k$ (see \eq~\ref{eq:Lambda_main}). \\
$\Mpn$ & Dimensionless mass ratio factor (see \eq~\ref{eq:Mpn}). \\
$\alpha(p.\mathrm{C1},k.\mathrm{C2};p)$ & `Sign quantity' $\alpha(p.\mathrm{C1},k.\mathrm{C2};p)$, for which $\alpha(p.\mathrm{C1},k.\mathrm{C2};p)=\pm1$; it ensures invariance of the Hamiltonian with respect to the choice of relative separation vectors (i.e., whether $\ve{r}_p$ points towards child 1 or child 2 in orbit $p$; see also Appendix A1 of Paper I). \\
$\{k.\mathrm{C}\}$ & Set of both children of binary $k$, i.e., which contains the components $\{k.\mathrm{C1}\}$ and $\{k.\mathrm{C2}\}$. \\
$M_{k.\mathrm{CS}(p)}$ & Mass of the sibling in orbit $k$ of the child in orbit $k$ that is connected to $p$. \\
$\Anm$ & Coefficients appearing in the Legendre polynomials (see \eq~\ref{eq:Adef}). \\
$\Bnmi$ & Dimensionless polynomial functions of $e_p$, defined implicitly by \eqs~(A125) and (A138a) of Paper I. In Table~\ref{table:AB}, we tabulate all values of $\Anm$ and $\Bnmi$ when $\Bnmi$ is nonzero, for $2\leq n \leq 5$. \\
$\Cnil$ & Dimensionless analytic functions of $e_p$ and $e_k$; they are defined implicitly in \eqs~(A133) and (A138b) of Paper I. They apply exclusively to pairwise double-averaged terms. \\
$\ve{\alpha}_k$, $\ve{\beta}_k$ & Regularised 4-vectors for orbit $k$, which are the KS analogs of $\ve{e}_i$ and $\ve{\j}_i$. \\
$\ve{u}_k$ & Regularised KS coordinate for orbit $k$. \\
$E_k$ & Generalised eccentric anomaly corresponding to orbit $k$. \\
$\ve{u}_k^\star$ & $\ve{u}_k^\star \equiv \mathrm{d} \ve{u}_i/\mathrm{d} E_i$. \\
$s_k$ & Regularised fictitious time for orbit $k$, defined according to the KS transformation $\mathrm{d} t/\mathrm{d} s_k = r_k$. \\
$\omega_k$ & KS frequency of orbit $k$ (see \eq~\ref{eq:KSomega}). \\
$V_k$ & Perturbing potential corresponding to orbit $k$. \\
$L^\mathrm{T}=L^\mathrm{T}(\ve{u}_k)$ & KS $L$-matrix for orbit $k$, depending on $\ve{u}_k$. \\
$\ve{P}_k$ & Perturbing acceleration for orbit $k$ (see \eq~\ref{eq:KSP}). \\
$\tau_k$ & KS time for orbit $k$ (\eq~\ref{eq:KStau}). \\
$P_{\mathrm{LK},pk}$ & Lidov-Kozai timescale associated with the orbit pair $(p,k$). \\
$P_{\mathrm{orb},k}$ & Orbital period of orbit $k$.\\
\bottomrule
\end{tabular}
\caption{Overview of the definitions of all important quantities used in this paper. }
\label{table:def}
\end{table}

\section{Methodology}
\label{sect:meth}
\subsection{Hamiltonian expansion}
\label{sect:meth:ham}

In Paper I, we derived the Hamiltonian for hierarchical systems composed of nested binary orbits, with arbitrary structure and number of bodies (in Paper II, we extended this formalism to include external perturbations). First, we expressed the Hamiltonian $H$ in terms of relative binary coordinates $\ve{r}_i$ (also known as Jacobi coordinates). Subsequently, we assumed that the system is hierarchical, and expanded $H$ in terms of the small ratios $r_i/r_j$ in the system, where $r_i$ and $r_j$ are the (instantaneous) separations of an inner and outer orbit, respectively. The result, which we repeat here, can be written in the form
\begin{align}
\label{eq:H}
H &=  H_\mathrm{Kep} + \sum_{n=2}^{\infty} S'_n,
\end{align}
where $H_\mathrm{Kep}$ is the Keplerian part of the Hamiltonian,
\begin{align}
\label{eq:Hkep}
H_\mathrm{Kep} = \sum_{k\in\mathrm{B}} \left [ \frac{1}{2} \frac{  M_{k.\mathrm{C1}} M_{k.\mathrm{C2}}}{M_k} \left (\dot{\ve{r}}_k \cdot \dot{\ve{r}}_k\right )
- \frac{GM_{k.\mathrm{C1}} M_{k.\mathrm{C2}}}{r_k} \right ],
\end{align}
and $S'_n$ represents the perturbing potential to order $n$. In \eq~(\ref{eq:Hkep}), $k \in \mathrm{B}$ represents a summation over all binaries in the system ($\mathrm{B}$ is the set of all binaries). The quantity $M_k$ is the total mass of binary $k$ (combining the mass of all bodies contained within it), and $M_{k.\mathrm{C1}}$ and $M_{k.\mathrm{C2}}$ are the masses of all bodies contained within child 1 and 2 of binary $k$, respectively (by definition, $M_k = M_{k.\mathrm{C1}} + M_{k.\mathrm{C2}}$). Dots denote derivatives with respect to time $t$, and $G$ is the gravitational constant. An overview of the definitions of all important quantities used in this paper is given in Table~\ref{table:def}.

The perturbing potential (which gives to long-term orbital changes) can be written as
\begin{align}
\label{eq:Spn}
S'_n = \sum_{j=2}^\infty S'_{n;j},
\end{align}
where $S'_{n;j}$ is the perturbing potential of order $n$ for interactions involving $j$ binaries. As shown in Paper I, the Hamiltonian is typically dominated by the pairwise terms ($j=2$). Also, only pairwise terms appear at the quadrupole order  ($n=2$), i.e., $S'_{2;j}=0$ for $j>2$. Terms involving 3 or more binaries ($j=3$) appear starting at octupole order ($n=3$), terms involving 4 or more binaries ($j=4$) appear starting at hexadecupole order ($n=4$), and so forth. The explicit expressions for the pairwise (any order $n$) and triplet terms (order $n=3$) are given below in \Ss~\ref{sect:meth:pair} and \ref{sect:meth:triplet}, respectively.

The equations of motion from the nonaveraged Hamiltonian for an orbit $i$ are generally given by
\begin{align}
\label{eq:EOM}
 &\ddot{\ve{r}}_{i} = -\frac{1}{\mu_i} \frac{\partial H}{\partial \ve{r}_i} =  - \frac{GM_i}{r_i^3} \ve{r}_i - \frac{1}{\mu_i} \sum_{n=2}^\infty \sum_{j=2}^{\infty} \frac{\partial S_{n;j}'}{\partial \ve{r}_i},
 \end{align}
where $\mu_i$ is the reduced mass of orbit $i$,
\begin{align}
\label{eq:mu}
\mu_i \equiv \frac{M_{i.\mathrm{C1}} M_{i.\mathrm{C2}}}{M_i},
\end{align}
and $\partial /\partial \ve{r}_i$ is shorthand notation for the gradient with respect to $\ve{r}_i$. Evidently, the first term after the second equality in \eq~(\ref{eq:EOM}) is the Keplerian acceleration, and the other terms represent perturbations. The Hamiltonian, \eq~(\ref{eq:H}), has been expanded in terms of separation ratios, but is exact in the limit that the system is hierarchical and that an infinite number of terms is included (both with respect to the expansion order $n$, and $j$-wise interactions). In this case, the resulting equations of motion are equivalent to directly solving the $N$-body equations of motion. 

Of course, in practice, the number of terms in the expansion of $H$ is limited, so $H$ is only approximate. Moreover, computing many high-order terms in the expansion of $H$ is computationally expensive, whereas the timescale of the integration is set by the shortest orbital period, which is similar to direct $N$-body integration. Therefore, instead of directly integrating the fully nonaveraged  equations of motion based on the expanded Hamiltonian, it is generally more efficient to use direct $N$-body integration, especially if special integration techniques are used such as algorithmic regularisation (e.g., \citealt{2006MNRAS.372..219M,2008AJ....135.2398M,2020MNRAS.492.4131R}). For completeness, we allow in \sm~the possibility to integrate all orbits directly, although in our tests, this method is inefficient compared to specialised direct $N$-body codes.  

The true advantage of the expansion of $H$ in terms of separation ratios becomes evident when at least some orbits in the system are averaged over. In particular, we consider cases when (some or all) inner orbits are averaged, whereas outer orbit(s) are integrated directly. Consequently, the typical integration timestep is increased from a fraction of the shortest orbital period, to a fraction of the timescale on which the inner orbit(s) change or the (typically much longer) timescale of the outer orbit(s), whichever is shorter. If the integration methods for all orbits are chosen judiciously, this hybrid method allows for accurate integration of the long-term evolution of the system, whereas being significantly faster compared to direct $N$-body integration. Evidently, when all orbits are averaged over, we recover the case assumed in Paper I. 

Below, we discuss this hybrid averaging approach in detail in the context of the pairwise terms at any order (\S~\ref{sect:meth:pair}), as well as the triplet terms at the octupole order (\S~\ref{sect:meth:triplet}).

\subsection{Pairwise interaction terms}
\label{sect:meth:pair}
\subsubsection{Nonaveraged}
\label{sect:meth:pair:un}
The pairwise perturbing potential was derived explicitly in Paper I; it is given by
\begin{align}
\label{eq:S_n_pair_gen}
S'_{n;2} &= \sum_{k \in \mathrm{B}} \sum_{\substack{p \in \mathrm{B} \\ p \in \{ k.\mathrm{C} \} }}  S'_{n;2}(p,k),
\end{align}
where
\begin{align}
\nonumber S'_{n;2}(p,k) &= (-1)^{n+1} \mu_p \, \Mpn \frac{GM_{k.\mathrm{CS}(p)}}{r_k} \sum_{m=0}^n \Anm \frac{ \left ( \ve{r}_p \cdot \ve{r}_k \right )^m r_p^{n-m} }{r_k^{n+m}}.
\end{align}
Here, we defined the dimensionless mass ratio factor for binary $p$,
\begin{align}
\label{eq:Mpn}
\Mpn \equiv \alpha(p.\mathrm{C1},k.\mathrm{C2};p)^n \frac{ M_{p.\mathrm{C2}}^{n-1} + (-1)^n M_{p.\mathrm{C1}}^{n-1}}{M_p^{n-1}}.
\end{align}
In addition, `$\{k.\mathrm{C}\}$' denotes the set of both children of binary $k$, i.e., which contains the components $\{k.\mathrm{C1}\}$ and $\{k.\mathrm{C2}\}$. We use the notation `$k.\mathrm{CS}(p)$' to denote the sibling in $k$ of the child in $k$ that is connected to $p$ (for example, in a triple, with $k$ representing the outer orbit and $p$ the inner orbit, $M_{k.\mathrm{CS}(p)}=m_3$, the tertiary mass). The `sign quantity' $\alpha(p.\mathrm{C1},k.\mathrm{C2};p)$, for which $\alpha(p.\mathrm{C1},k.\mathrm{C2};p)=\pm1$, ensures invariance of the Hamiltonian with respect to the choice of relative separation vectors (i.e., whether $\ve{r}_p$ points towards child 1 or child 2 in orbit $p$; see also Appendix A1 of Paper I). 
The dimensionless quantities $\Anm$ are coefficients appearing in the Legendre polynomials. They can be determined from the Rodrigues formula, i.e.,
\begin{align}
\label{eq:Adef}
\sum_{m=0}^n \Anm x^m = \frac{1}{2^n n!} \frac{\mathrm{d}^n }{\mathrm{d} x^n} \left [ \left (x^2 -1 \right )^n \right ].
\end{align}

The contribution from the pairwise terms in the Hamiltonian to the equations of motion (to order $n$) is given by
\begin{align}
\label{eq:EOM_rp_pair}
\nonumber &\ddot{\ve{r}}_{p}= - \frac{1}{\mu_p} \frac{ \partial S'_{n;2}(p,k)}{\partial \ve{r}_p} = (-1)^n \Mpn GM_{k.\mathrm{CS}(p)}  \left ( \frac{r_p}{r_k} \right )^n \\
&\quad \times \sum_{m=0}^n \Anm \frac{ \left ( \ve{r}_p \cdot \ve{r}_k \right )^{m-1}}{r_p^{m+2} r_k^{m+1}} \biggl [ m r_p^2 \ve{r}_k 
+ (n-m) \left ( \ve{r}_p \cdot \ve{r}_k \right ) \ve{r}_p \biggl ]
\end{align}
for the inner orbit in the pair $(p,k)$; for the outer orbit,
\begin{align}
\label{eq:EOM_rk_pair}
\nonumber &\ddot{\ve{r}}_{k}= - \frac{1}{\mu_k} \frac{ \partial S'_{n;2}(p,k)}{\partial \ve{r}_k} = (-1)^n \frac{\mu_p}{\mu_k} \Mpn GM_{k.\mathrm{CS}(p)}  \left ( \frac{r_p}{r_k} \right )^n \\
&\quad \times \sum_{m=0}^n \Anm \frac{ \left (\ve{r}_p\cdot \ve{r}_k\right)^{m-1}}{r_p^{m}r_k^{m+3}} \left [ m\, r_k^2 \ve{r}_p - (n+m+1) \left ( \ve{r}_p \cdot \ve{r}_k \right ) \ve{r}_k \right ].
\end{align}

\subsubsection{Single averaged}
In the single-averaging approach for pairwise interactions, we average over the inner orbit of each pair, but integrate the outer orbit directly. Let $\ve{e}_p$ and $\ve{\j}_p$, with $\j_p = \sqrt{1-e_p^2}$, denote the eccentricity and normalised angular-momentum vectors of orbit $p$, respectively, and let $a_p$ denote the semimajor axis. The resulting perturbing potential is given by (see Paper I, \S~A5.3)
\begin{align}
\label{eq:S_n_pair_gen_SA}
\nonumber &\left \langle S'_{n;2}(p,k) \right \rangle_p =  (-1)^{n+1} \mu_p  \Mpn \frac{GM_{k.\mathrm{CS}(p)}}{r_k} \left (\frac{a_p}{r_k} \right )^n  \\
&\quad \times \sum_{m=0}^n \Anm \sum_{\substack{i_1,i_2 \in \, \mathbb{N}^0 \\ i_1+i_2 \leq m}} \Bnmi  \frac{ \left ( \ve{e}_p \cdot \ve{r}_k \right )^{i_1} \left ( \ve{\j}_p \cdot \ve{r}_k \right )^{i_2} }{r_k^{i_1+i_2}}.
\end{align}
Here, $\mathbb{N}^0$ are the natural numbers plus zero, and we introduced the (completely analytic) functions $\Bnmi$, which are polynomial functions of $e_p$ (i.e., the eccentricity of the orbit that has been averaged over) and are defined implicitly by \eqs~(A125) and (A138a) of Paper I. In Table~\ref{table:AB}, we tabulate all values of $\Anm$ and $\Bnmi$ when $\Bnmi$ is nonzero, for $2\leq n \leq 5$.

\begin{table}
\begin{tabular}{cccccc}
\toprule
$n$ & $m$ & $i_1$ & $i_2$ & $\Anm$ & $\Bnmie$ \\
\midrule
 2 & 0 & 0 & 0 & $-\frac{1}{2}$ & $\frac{3}{2}e^2+1$ \\
 2 & 2 & 0 & 0 & $\frac{3}{2}$ & $\frac{1}{2}
   \left(1-e^2\right)$ \\
 2 & 2 & 0 & 2 &$ \frac{3}{2} $&$ -\frac{1}{2}$ \\
 2 & 2 & 2 & 0 & $\frac{3}{2} $& $\frac{5}{2} $\\
 3 & 1 & 1 & 0 & $-\frac{3}{2}$ & $-\frac{5}{8} \left(3
   e^2+4\right)$ \\
 3 & 3 & 1 & 0 & $\frac{5}{2}$ & $\frac{15}{8}
   \left(e^2-1\right)$ \\
 3 & 3 & 1 & 2 & $\frac{5}{2}$ & $\frac{15}{8} $\\
 3 & 3 & 3 & 0 & $\frac{5}{2}$ & $-\frac{35}{8} $\\
 4 & 0 & 0 & 0 & $\frac{3}{8} $& $\frac{15}{8}e^4+5
   e^2+1$ \\
 4 & 2 & 0 & 0 & $-\frac{15}{4}$ & $\frac{1}{8} \left(-3
   e^4-e^2+4\right)$ \\
 4 & 2 & 0 & 2 & $-\frac{15}{4}$ &$ \frac{1}{8} \left(-3
   e^2-4\right)$ \\
 4 & 2 & 2 & 0 & $-\frac{15}{4}$ &$ \frac{21}{8}
   \left(e^2+2\right) $\\
 4 & 4 & 0 & 0 & $\frac{35}{8}$&$ \frac{3}{8}
   \left(e^2-1\right)^2$ \\
 4 & 4 & 0 & 2 & $\frac{35}{8}$ & $\frac{3}{4}
   \left(e^2-1\right) $\\
 4 & 4 & 0 & 4 & $\frac{35}{8}$ & $\frac{3}{8}$\\
 4 & 4 & 2 & 0 & $\frac{35}{8}$ & $ -\frac{21}{4}
   \left(e^2-1\right) $\\
 4 & 4 & 2 & 2 &$ \frac{35}{8}$ & $-\frac{21}{4}$ \\
 4 & 4 & 4 & 0 & $\frac{35}{8}$ & $\frac{63}{8} $\\
 5 & 1 & 1 & 0 & $\frac{15}{8}$ &$ -\frac{7}{16} \left(5
   e^4+20 e^2+8\right) $\\
 5 & 3 & 1 & 0 & $-\frac{35}{4}$ &$ \frac{21}{16}
   \left(e^4+e^2-2\right) $\\
 5 & 3 & 1 & 2 & $-\frac{35}{4}$ & $\frac{21}{16}
   \left(e^2+2\right) $\\
 5 & 3 & 3 & 0 &$ -\frac{35}{4}$ &$ -\frac{21}{16} \left(3
   e^2+8\right) $\\
 5 & 5 & 1 & 0 & $\frac{63}{8}$ & $-\frac{35}{16}
   \left(e^2-1\right)^2$ \\
 5 & 5 & 1 & 2 & $\frac{63}{8}$ & $-\frac{35}{8}
   \left(e^2-1\right)$ \\
 5 & 5 & 1 & 4 & $\frac{63}{8}$ & $-\frac{35}{16}$ \\
 5 & 5 & 3 & 0 & $\frac{63}{8}$ & $\frac{105}{8}
   \left(e^2-1\right) $\\
 5 & 5 & 3 & 2 &$ \frac{63}{8}$ &$ \frac{105}{8}$ \\
 5 & 5 & 5 & 0 & $\frac{63}{8}$ & $-\frac{231}{16} $\\
\bottomrule
\end{tabular}
\caption{All values of $\Anm$ (\eq~\ref{eq:Adef}) and $\Bnmie$ (\eqs~A125 and A138a of Paper I) when $\Bnmie$ is nonzero, and for $2\leq n \leq 5$. }
\label{table:AB}
\end{table}

The inner-averaged pairwise perturbing potential gives rise to secular changes of the inner orbit given by the Milankovitch equations (\citealt{milankovitch_39}; e.g., \citealt{1961JGR....66.2797M,1963PCPS...59..669A,1964RSPSA.280...97A,2005MNRAS.364.1222B,2009AJ....137.3706T}; see \citealt{2014CeMDA.118..197R} for an overview). Specifically, for an orbit $i$,
\begin{subequations}
\label{eq:EOM_Mil}
\begin{align}
\label{eq:EOM_Mil:j}
\frac{\mathrm{d} \ve{\j}_i}{\mathrm{d} t} &= - \frac{1}{\Lambda_i} \left [ \, \ve{\j}_i \times \frac{\partial \left \langle S'_{n;2}(p,k) \right \rangle_p}{\partial \ve{\j}_i} + \ve{e}_i \times  \frac{\partial \left \langle S'_{n;2}(p,k) \right \rangle_p}{\partial \ve{e}_i} \, \right ]; \\
\label{eq:EOM_Mil:e}
\frac{\mathrm{d} \ve{e}_i}{\mathrm{d} t} &= - \frac{1}{\Lambda_i} \left [ \, \ve{e}_i \times \frac{\partial \left \langle S'_{n;2}(p,k) \right \rangle_p}{\partial \ve{\j}_i} + \ve{\j}_i \times  \frac{\partial \left \langle S'_{n;2}(p,k) \right \rangle_p}{\partial \ve{e}_i} \, \right ].
\end{align}
\end{subequations}
Here,
\begin{align}
\label{eq:Lambda_main}
\Lambda_i = \mu_i \, \sqrt{G M_i a_i}
\end{align}
is the circular angular momentum of orbit $i$. The gradients with respect to $\ve{e}_i$ and $\ve{\j}_i$ are given explicitly by
\begin{subequations}
\begin{align}
\nonumber &\frac{\partial  \left \langle S'_{n;2}(p,k) \right \rangle_p}{\partial \ve{e}_p} = (-1)^{n+1} \mu_p  \Mpn GM_{k.\mathrm{CS}(p)} \\
\nonumber &\quad \times \sum_{m=0}^n \Anm a_p^n \sum_{\substack{i_1,i_2 \in \, \mathbb{N}^0 \\ i_1+i_2 \leq m}} \frac{\left(\ve{\j}_p \cdot \ve{r}_k \right )^{i_2}}{r_k^{n+1+i_1+i_2}} \biggl [ \frac{\partial \Bnmi}{\partial e_p} \left ( \ve{e}_p \cdot \ve{r}_k \right )^{i_1} \unit{e}_p \\
&\qquad + i_1 \Bnmi \left ( \ve{e}_p \cdot \ve{r}_k \right )^{i_1-1} \ve{r}_k \biggl ]; \\
\nonumber &\frac{\partial \left \langle S'_{n;2}(p,k) \right \rangle_p}{\partial \ve{\j}_p} = (-1)^{n+1} \mu_p \Mpn GM_{k.\mathrm{CS}(p)} \\
\nonumber &\quad \times \sum_{m=0}^n \Anm a_p^n \sum_{\substack{i_1,i_2 \in \, \mathbb{N}^0 \\ i_1+i_2 \leq m}} \Bnmi \frac{\left(\ve{e}_p \cdot \ve{r}_k \right )^{i_1}}{r_k^{n+1+i_1+i_2}} i_2 \left ( \ve{\j}_p \cdot \ve{r}_k \right )^{i_2-1} \ve{r}_k
\end{align}
\end{subequations}
(here, hats denote unit vectors). The corresponding contributions to the equations of motion for the (nonaveraged) outer orbit are
\begin{align}
\nonumber &\ddot{\ve{r}}_{k}= - \frac{1}{\mu_k} \frac{ \partial \left \langle S'_{n;2}(p,k) \right \rangle_p}{\partial \ve{r}_k} = (-1)^n \frac{\mu_p}{\mu_k} \Mpn GM_{k.\mathrm{CS}(p)}  \\
\nonumber &\quad \times\sum_{m=0}^n \Anm \left ( \frac{a_p}{r_k} \right )^n  \sum_{\substack{i_1,i_2 \in \, \mathbb{N}^0 \\ i_1+i_2 \leq m}} \Bnmi  \frac{\left ( \ve{e}_p \cdot \ve{r}_k \right )^{i_1-1} \left ( \ve{\j}_p \cdot \ve{r}_k \right )^{i_2-1}}{r_k^{i_1+i_2+3}}\\
\nonumber &\quad \times \biggl [ i_1 \vdot{\ve{\j}_p}{\ve{r}_k} r_k^2 \ve{e}_p + i_2 \vdot{\ve{e}_p}{\ve{r}_k} r_k^2 \ve{\j}_p \\
&\qquad - (n+1+i_1+i_2) \vdot{\ve{e}_p}{\ve{r}_k} \vdot{\ve{\j}_p}{\ve{r}_k} \ve{r}_k \biggl ].
\end{align}

\subsubsection{Double averaged}
In this approach for pairwise interactions, both orbits are averaged over. The perturbing potential is given by (see Paper I, \S~A5.3)
\begin{align}
\label{eq:S_n_pair_gen_final}
\nonumber &\left \langle S'_{n;2}(p,k) \right \rangle_{p,k} =  (-1)^{n+1} \mu_p \Mpn \frac{GM_{k.\mathrm{CS}(p)}}{a_k}  \left ( \frac{a_p}{a_k} \right )^n  \\
\nonumber &\quad \times\frac{1}{j_k^{2n-1}} \sum_{m=0}^n \Anm \sum_{\substack{i_1,i_2 \in \, \mathbb{N}^0 \\ i_1+i_2 \leq m}} \Bnmi \\
\nonumber &\quad \times \sum_{\substack{l_1,l_2,l_3,l_4 \in \, \mathbb{N}^0 \\ l_1+l_3 \leq i_1 \\ l_2 + l_4\leq i_2}}  \, \Cnil \left ( \ve{e}_p\cdot \ve{e}_k \right )^{l_1} \left ( \ve{j}_p\cdot \ve{j}_k \right )^{l_2} \\
&\qquad \times \left ( \ve{e}_p\cdot \ve{j}_k \right )^{l_3} \left ( \ve{e}_k \cdot \ve{j}_p \right )^{l_4}.
\end{align}
Here, $\Cnil$ are analytic functions of $e_p$ and $e_k$; they are defined implicitly in \eqs~(A133) and (A138b) of Paper I. The equations of motion for both orbits $p$ and $k$ are straightforwardly derived from the Milankovitch equations (\eqs~\ref{eq:EOM_Mil}). Since the double-averaging approximation was already the focus of Paper I, we here do not show the explicit expressions of the equations of motion.

\subsection{Triplet interaction terms}
\label{sect:meth:triplet}
\subsubsection{Nonaveraged}
\label{sect:meth:triplet:un}
In Paper I, we showed that `cross' terms, i.e., terms that individually involve more than two binaries, start appearing in the Hamiltonian at sufficiently high expansion orders. The simplest cross term is the `triplet' term depending on three binaries, which appears starting at octupole order in systems that have at three orbits on different levels. For example, the triplet term appears at octupole order in 3+1 quadruple systems, which have three different levels, but not in 2+2 quadruple systems, which have only two different levels (see, also, \citealt{2015MNRAS.449.4221H}). 

Generally, cross terms are much smaller in magnitude compared to pairwise terms, since they involve separation ratios of nonadjacent orbits which are, evidently, smaller than separation ratios of adjacent orbits. Moreover, cross terms are generally complicated, especially at high expansion orders. Nevertheless, to be able to investigate their importance in numerical integrations, we here include the lowest-order cross term, i.e., the octupole-order triplet term. The latter was derived in Paper I, and is given by
\begin{align}
&S_{3;3}' = \sum_{k \in \mathrm{B}} \sum_{\substack{p \in \mathrm{B} \\ p\in\{k.\mathrm{C} \}}} \sum_{\substack{u \in \mathrm{B} \\ u\in\{k.\mathrm{C} \} \\ p \in \{u.\mathrm{C} \} }} S_{3;3}'(p,u,k),
\end{align}
where
\begin{align}
\nonumber &S_{3;3}'(p,u,k) = \frac{3}{2} \mu_p \alpha(p,k.\mathrm{CS}(p);k) \frac{G M_{k.\mathrm{CS}(p)}}{r_k} \\
\nonumber &\quad \times \frac{\alpha(p,k.\mathrm{CS}(p);u) M_{u.\mathrm{CS}(p)}}{M_u} \left ( \frac{r_p}{r_k} \right )^2 \left ( \frac{r_u}{r_k} \right ) \\
&\quad \times \left [ 5 \left ( \unit{r}_p \cdot \unit{r}_k \right )^2 \left ( \unit{r}_u \cdot \unit{r}_k \right ) - 2 \left ( \unit{r}_p \cdot \unit{r}_k \right ) \left ( \unit{r}_p \cdot \unit{r}_u \right ) - \left ( \unit{r}_u \cdot \unit{r}_k \right ) \right ].
\label{eq:H_triplet}
\end{align}
Note that, since $r_p<r_u<r_k$ in \eq~(\ref{eq:H_triplet}), $(r_p/r_k)^2 \ll (r_p/r_u)^2$ and $(r_p/r_k)^2 \ll (r_u/r_k)^2$. Therefore, $S_{3;3}'$ is much smaller than the corresponding $S'_{3;2}$ applied to the $(p,u)$ and $(u,k)$ pairs, respectively. 

The equations of motion resulting from the (nonaveraged) triplet Hamiltonian are given in Appendix~\ref{app:H_triplet:un}. 

\subsubsection{Inner averaged}
Averaging the octupole-order triplet Hamiltonian over the inner orbit gives
\begin{align}
\nonumber &\left \langle S_{3;3}'(p,u,k) \right \rangle_p = \frac{3}{4} \mu_p \alpha(p,k.\mathrm{CS}(p);k) \,G M_{k.\mathrm{CS}(p)} \\
\nonumber &\quad \times \frac{\alpha(p,k.\mathrm{CS}(p);u) M_{u.\mathrm{CS}(p)}}{M_u} \frac{a_p^2}{r_k^7} \Biggl [ \vdot{\ve{r}_u}{\ve{r}_k} \left \{ \left(1-e_p^2 \right ) r_k^2 \right. \\
\nonumber &\qquad \left. + 5 \vdot{\ve{e}_p}{\ve{r}_k}^2 - \vdot{\ve{\j}_p}{\ve{r}_k}^2 \right \} -2 r_k^2 \left \{ \left(1-e_p^2 \right )\vdot{\ve{r}_u}{\ve{r}_k} \right. \\
\nonumber &\qquad \left. + 5 \vdot{\ve{e}_p}{\ve{r}_u} \vdot{\ve{e}_p}{\ve{r}_k} - \vdot{\ve{\j}_p}{\ve{r}_u}\vdot{\ve{\j}_p}{\ve{r}_k} \right \} \\
&\qquad - r_k^2 \vdot{\ve{r}_u}{\ve{r}_k} \left(2+3e_p^2 \right ) \Biggl ].
\label{eq:H_triplet_A}
\end{align}
The corresponding equations of motion are given in Appendix~\ref{app:H_triplet:A}.

\subsubsection{Double averaged}
Here, we average over the inner and intermediate orbits, $p$ and $u$. The result is
\begin{align}
\nonumber &\left \langle S_{3;3}'(p,u,k) \right \rangle_{p,u} = -\frac{9}{8} \mu_p \alpha(p,k.\mathrm{CS}(p);k) \,G M_{k.\mathrm{CS}(p)} \\
\nonumber &\quad \times \frac{\alpha(p,k.\mathrm{CS}(p);u) M_{u.\mathrm{CS}(p)}}{M_u} \frac{a_p^2a_u}{r_k^7} \Biggl [ \vdot{\ve{e}_u}{\ve{r}_k} \left \{ \left(1-e_p^2 \right ) r_k^2 \right. \\
\nonumber &\qquad \left. + 5 \vdot{\ve{e}_p}{\ve{r}_k}^2 - \vdot{\ve{\j}_p}{\ve{r}_k}^2 \right \} -2 r_k^2 \left \{ \left(1-e_p^2 \right )\vdot{\ve{e}_u}{\ve{r}_k} \right. \\
\nonumber &\qquad \left. + 5 \vdot{\ve{e}_p}{\ve{e}_u} \vdot{\ve{e}_p}{\ve{r}_k} - \vdot{\ve{\j}_p}{\ve{e}_u}\vdot{\ve{\j}_p}{\ve{r}_k} \right \} \\
&\qquad - r_k^2 \vdot{\ve{e}_u}{\ve{r}_k} \left(2+3e_p^2 \right ) \Biggl ].
\label{eq:H_triplet_AB}
\end{align}
The corresponding equations of motion are given in Appendix~\ref{app:H_triplet:AB}.

\subsubsection{Triple averaged}
Lastly, averaging over all orbits, $p$, $u$, and $k$, gives
\begin{align}
\nonumber &\left \langle S_{3;3}'(p,u,k) \right \rangle_{p,u,k} = \mu_p \alpha(p,k.\mathrm{CS}(p);k) \frac{\alpha(p,k.\mathrm{CS}(p);u) M_{u.\mathrm{CS}(p)}}{M_u} \\
\nonumber &\quad \times \frac{G M_{k.\mathrm{CS}(p)}}{a_k} \left ( \frac{a_p}{a_k} \right )^2 \left ( \frac{a_u}{a_k} \right ) \frac{9}{32 j_k^7} \Biggl [10 (\ve{e}_p \cdot \ve{e}_u) (\ve{e}_p \cdot \ve{e}_k) j_k^2  \\
\nonumber &\qquad  - 50 (\ve{e}_p \cdot \ve{e}_k) (\ve{e}_p \cdot \ve{j}_k) (\ve{e}_u \cdot \ve{j}_k) - 2 (\ve{e}_k \cdot \ve{j}_p) (\ve{e}_u \cdot \ve{j}_p) j_k^2 \\
\nonumber &\qquad + 10 (\ve{e}_u \cdot \ve{j}_k) (\ve{e}_k \cdot \ve{j}_p) (\ve{j}_p \cdot \ve{j}_k)  \\
&\quad \quad  - (\ve{e}_u \cdot \ve{e}_k) \left \{ \left(1-6e_p^2\right)j_k^2 + 25 (\ve{e}_p \cdot \ve{j}_k)^2 - 5 (\ve{j}_p \cdot \ve{j}_k)^2 \right \} \Biggl ].
\end{align}
Similarly to the discussion of the fully-averaged pairwise terms above, we here do not explicitly give the corresponding equations of motion based on the fully-averaged triplet terms since the fully-averaged case was the focus of Paper I.

\subsection{Direct integration methods}
\label{sect:meth:dir}
Within \sm, the averaged orbits are integrated by including the corresponding equations of motion for the orbital vectors, $\dot{\ve{e}}_i$ and $\dot{\ve{\j}}_i$ (see \eq~\ref{eq:EOM_Mil}), in the set of first-order ordinary differential equations (ODEs). We adopt two techniques to propagate nonaveraged orbits. The most straightforward approach is to include both $\ve{r}_i$ and $\ve{v}_i \equiv \dot{\ve{r}}_i$ in the set of ODEs (six variables per orbit), with their time derivatives simply given by 
\begin{align}
\left \{
\begin{array}{cc}
\displaystyle \frac{\mathrm{d}\ve{r}_i}{\mathrm{d} t} &= \ve{v}_i; \\
\\
\displaystyle \frac{\mathrm{d}\ve{v}_i}{\mathrm{d} t} &= \ddot{\ve{r}}_i, 
\end{array} \right.
\end{align}
where $\ddot{\ve{r}}_i$ is determined by \eq~(\ref{eq:EOM}). This method, although straightforward, suffers from the fact that $\ve{r}_i$ and $\ve{v}_i$ still need to be propagated in the case without perturbations (i.e., $S'_{n;j}=0$ in \eq~\ref{eq:EOM} so $\ddot{\ve{r}}_i = - GM_i/r_i^3 \, \ve{r}_i$), whereas the orbit is static. This is prone to cause numerical errors. 

Therefore, we also implemented an alternative integration approach based on the regularised Kustaanheimo-Stiefel (KS) equations of motion in element form (\citealt{1971lrcm.book.....S}; see also, e.g., \citealt{2017rom..book.....R}). These equations are formulated in terms of the regularised 4-vectors $\ve{\alpha}_i$ and $\ve{\beta}_i$, which are the KS analogs of $\ve{e}_i$ and $\ve{\j}_i$. The relation between $\ve{\alpha}_i$ and $\ve{\beta}_i$ and the standard regularised KS coordinates $\ve{u}_i$ is
\begin{subequations}
\label{eq:KS_uab}
\begin{align}
\ve{u}_i &= \ve{\alpha}_i \cos \frac{E_i}{2} + \ve{\beta}_i \sin \frac{E_i}{2}; \\
\ve{u}_i^\star &= -\frac{1}{2} \ve{\alpha}_i \sin \frac{E_i}{2} + \frac{1}{2} \ve{\beta}_i \cos \frac{E_i}{2},
\end{align}
\end{subequations}
where $E_i$ is the generalised eccentric anomaly corresponding to orbit $i$, and $\ve{u}_i^\star \equiv \mathrm{d} \ve{u}_i/\mathrm{d} E_i$. The generalised eccentric anomaly is directly related to the fictitious time $s_i$ (which is defined according to the usual KS transformation $\mathrm{d} t/\mathrm{d} s_i = r_i$) via
\begin{align}
E_i = 2 \omega_i s_i,
\end{align}
where $\omega_i$ is the KS frequency, defined according to
\begin{align}
\label{eq:KSomega}
2 \omega_i^2 = \frac{GM_i}{r_i} - \frac{1}{2} \dot{\ve{r}}_i^2 - V_i,
\end{align}
with $V_i$ the perturbing potential corresponding to orbit $i$. The element equations (setting $\partial V_i/\partial t=0$, as appropriate in our case with a conservative potential) are then
\begin{subequations}
\label{eq:KS_E}
\begin{align}
\label{eq:KS_E_omega}
\frac{\mathrm{d} \omega_i}{\mathrm{d} E_i} &= - \frac{1}{2\omega_i} \left(\ve{u}_i^\star \cdot L^\mathrm{T} \ve{P}_i\right ); \\
\label{eq:KS_E_tau}
\nonumber \frac{\mathrm{d} \tau_i}{\mathrm{d} E_i} &= \frac{1}{8\omega_i^3} \left ( G^2M_i^2 - 2 r_i V_i \right ) - \frac{r_i}{16 \omega_i^3} \left [ \left ( \ve{u}_i \cdot \frac{\partial V_i}{\partial \ve{u}_i} \right ) - 2 L^\mathrm{T} \ve{P}_i \right ] \\
&\quad - \frac{2}{\omega_i^2} \frac{\mathrm{d} \omega_i}{\mathrm{d} E_i} \left ( \ve{u}_i \cdot \ve{u}_i^\star \right ); \\
\frac{\mathrm{d} \ve{\alpha}_i}{\mathrm{d} E_i} &= \Biggl \{ \frac{1}{2 \omega_i^2} \left [ \frac{V_i}{2} + \frac{r_i}{4} \left ( \frac{\partial V_i}{\partial \ve{u}_i} - 2 L^\mathrm{T}\ve{P}_i \right ) \right ] + \frac{2}{\omega_i} \frac{\mathrm{d} \omega_i}{\mathrm{d} E_i} \ve{u}_i^\star \Biggl \} \sin \frac{E_i}{2}; \\
\frac{\mathrm{d} \ve{\beta}_i}{\mathrm{d} E_i} &= -\Biggl \{ \frac{1}{2 \omega_i^2} \left [ \frac{V_i}{2} + \frac{r_i}{4} \left ( \frac{\partial V_i}{\partial \ve{u}_i} - 2 L^\mathrm{T}\ve{P}_i \right ) \right ] + \frac{2}{\omega_i} \frac{\mathrm{d} \omega_i}{\mathrm{d} E_i} \ve{u}_i^\star \Biggl \} \cos \frac{E_i}{2}.
\end{align}
\end{subequations}
Here, $L^\mathrm{T}=L^\mathrm{T}(\ve{u}_i)$ denotes the KS $L$-matrix (which depends on $\ve{u}_i$). $\ve{P}_i$ is the perturbing acceleration (the acceleration minus the Keplerian part), i.e., 
\begin{align}
\label{eq:KSP}
\ve{P}_i = \ddot{\ve{r}}_i + \frac{GM_i}{r_i^3} \ve{r}_i.
\end{align}
Furthermore, $\tau_i$ is related to physical time according to
\begin{align}
\label{eq:KStau}
t = \tau_i - \frac{1}{\omega_i} \left ( \ve{u}_i \cdot \ve{u}_i^\star \right ).
\end{align}

The main advantage of the KS equations in element form is that, in the absence of perturbations, $V_i=0$ and $\ve{P}_i=\ve{0}$, and the equations of motion simply state that $\ve{\alpha}_i$ and $\ve{\beta}_i$ are constants of the motion. This implies that no numerical errors are made in the orbit propagation in the absence of perturbations, which is numerically advantageous. 

The original KS element equations of motion are formulated in terms of the independent parameter $E$, the generalised eccentric anomaly. The equations apply to a single perturbed orbit (hence, there is no need to use the subscript $i$). In our case, however, there can be an arbitrary number of orbits that need to be propagated. This requires a formulation of the equations of motion in terms of a new, global parameter which depends on properties of all orbits, instead of the traditional KS transformation, $\mathrm{d} t/\mathrm{d} s = r$, for a single orbit. To our knowledge, such a formulation has not yet been produced, and its development is beyond the scope of this work. 

The implication, however, is that in our case, the KS element equations cannot be formulated in terms of a single $E$, but need to be formulated in terms of the global time $t$. This has the disadvantage that the relation between $E_i$ and the physical time for each orbit,
\begin{align}
\label{eq:KS_dEdt}
\frac{\mathrm{d} E_i}{\mathrm{d} t} = \frac{2\omega_i}{r_i},
\end{align}
contains a singularity with respect to $r_i$. Of course, it was the original purpose of the KS transformation to eliminate this singularity. However, in our case, in typical use cases we do not expect to integrate directly over the {\it innermost} orbits in the system. This implies that, for a given orbit $i$ that contains inner orbits, $r_i$ cannot approach zero in order, to retain the hierarchy (if $r_i \rightarrow0$ for an outer orbit $i$, the system would become dynamically unstable). Therefore, we do not expect that this limitation poses a major problem in our case. However, by using the KS element equations of motion, we still retain the advantage of propagating the orbits exactly in the absence of perturbations. 

In our case of a conservative system, one can choose in \eqs~(\ref{eq:KS_E}) between (1) using the perturbing potential $V_i$, or (2) the perturbing acceleration $\ve{P}_i$. In the former case, the perturbing potential is nonzero and determined by the Hamiltonian, whereas $\ve{P}_i=\ve{0}$. In the latter case, one can set $V_i=0$ whereas $\ve{P}_i \neq 0$. As discussed in \citet{1971lrcm.book.....S}, use of the perturbing potential, method (1), is generally preferred because it implies that $\omega_i$ is constant (see \eq~\ref{eq:KS_E_omega}), which is numerically an advantageous property. We implemented both methods (see \S~\ref{sect:use:ex}), but, in practice, found little to no differences in performance or accuracy of the integration.

In summary, in \sm, we implemented the KS element equations using physical time $t$ as the independent variable, and assuming either the perturbing potential or perturbing acceleration approaches. In the former case ($\ve{P}_i=\ve{0}$), 
\begin{subequations}
\begin{align}
\frac{\mathrm{d} \ve{\alpha}_i}{\mathrm{d} t} &= \frac{1}{\omega_i r_i} \left ( \frac{V_i}{2} \ve{u}_i + \frac{r_i}{4} \frac{\partial V_i}{\partial \ve{u}_i} \right ) \sin \frac{E_i}{2}; \\
\frac{\mathrm{d} \ve{\beta}_i}{\mathrm{d} t} &= -\frac{1}{\omega_i r_i} \left ( \frac{V_i}{2} \ve{u}_i + \frac{r_i}{4} \frac{\partial V_i}{\partial \ve{u}_i} \right ) \cos \frac{E_i}{2}.
\end{align}
\end{subequations}
In the latter case ($V_i=0$), 
\begin{subequations}
\begin{align}
\frac{\mathrm{d} \ve{\alpha}_i}{\mathrm{d} t} &=\Biggl [ - \frac{1}{2\omega_i} L^\mathrm{T} \ve{P}_i + \frac{4}{r_i} \frac{\mathrm{d} \omega_i}{\mathrm{d} E_i} \ve{u}_i^\star \Biggl ] \sin \frac{E_i}{2}; \\
\frac{\mathrm{d} \ve{\beta}_i}{\mathrm{d} t} &=-\Biggl [ - \frac{1}{2\omega_i} L^\mathrm{T} \ve{P}_i + \frac{4}{r_i} \frac{\mathrm{d} \omega_i}{\mathrm{d} E_i} \ve{u}_i^\star \Biggl ] \cos \frac{E_i}{2},
\end{align}
\end{subequations}
with $\mathrm{d} \omega_i/\mathrm{d} E_i$ given by \eq~(\ref{eq:KS_E_omega}). In both cases, $E_i$ is evolved according to \eq~(\ref{eq:KS_dEdt}). We reiterate that the apparent singularities in these equations with respect to $r_i$ do not pose a major problem in practice, since $r_i$ is never expected to be close to zero. 

Without loss of generality, we set the initial $E_i=0$, such that the initial $\ve{\alpha}_i$ and $\ve{\beta}_i$ are directly given by $\ve{\alpha}_i = \ve{u}_i$, and $\ve{\beta}_i = 2 \ve{u}_i^\star$ (see \eq~\ref{eq:KS_uab}). We refer to \citeauthor{1971lrcm.book.....S} (\citeyear{1971lrcm.book.....S}; Section 19) for the explicit transformation relations between the (nonregularised) Cartesian coordinates $\ve{r}_i$ and $\ve{v}_i$, and the regularised coordinates $\ve{u}_i$ and $\ve{u}_i^\star$.

\subsection{Orbit-averaging corrections}
\label{sect:meth:cor}
As alluded to in the Introduction, we also implemented orbit-averaging corrections within \sm. Such corrections take into account the response of an inner orbit to the outer orbit during the outer orbital timescale, which can accumulate over time and affect the long-term secular evolution. Analytical orbit-averaging corrections have been derived for hierarchical triple systems to the quadrupole expansion order in the test particle limit (where one of the bodies in the inner binary is massless, such that the outer orbit is static), in orbital vector form \citep{2016MNRAS.458.3060L}. Additionally, \citet{2018MNRAS.481.4602L} derived corrections to any expansion order in the test particle limit using orbital elements, and \citet{2019MNRAS.490.4756L} derived similar corrections also taking into account changes on the inner orbital timescale. However, to our knowledge, no extension has yet been made to the non-test-particle limit, and higher-multiplicity hierarchical systems. 

Such investigation is beyond the scope of this work. Instead, we here implemented in \sm~the orbit-averaging correction terms of \citet{2016MNRAS.458.3060L} in vector form which strictly apply to triples and in the test-particle limit. We included these terms, which are formulated as additional terms to $\dot{\ve{e}}_i$ and $\dot{\ve{\j}}_i$ for an inner orbit; the outer orbit is unaffected as the test-particle approximation is assumed. We refer to \citet{2016MNRAS.458.3060L} for the explicit expressions of the correction terms to $\dot{\ve{e}}_i$ and $\dot{\ve{\j}}_i$. Also, we note that there are caveats to this approach (see the discussion in \S~\ref{sect:discussion:cor}). 

We make the technical remark that \citet{2016MNRAS.458.3060L} assumed that the perturber orbital plane is aligned with the $z$-axis, and with the periapsis aligned along the $x$-direction. However, in \sm, the outer orbit can be aligned along an arbitrary direction. Therefore, in the practical implementation, we project $\ve{e}_\mathrm{in}$ and $\ve{\j}_\mathrm{in}$ of the inner orbit onto the eccentricity and angular-momentum vectors of the outer orbit, $\ve{e}_\mathrm{out}$ and $\ve{\j}_\mathrm{out}$, respectively. After computing the equations of motion for the projected $\ve{e}_\mathrm{in}$ and $\ve{\j}_\mathrm{in}$, the latter quantities are transformed back into the original frame used in \sm.

\newpage
\section{Code usage}
\label{sect:use}

\begin{lstlisting}[caption={Illustration in the \textsc{Python} language of how to use the \sm~code to integrate a triple in the single-averaging approximation. },label={code:use},language=Python]
from secularmultiple import SecularMultiple,Particle,Tools

### Generate particles ###
particles = Tools.create_nested_multiple(3, [m1,m2,m3],[a1,a2],[e1,e2],[i1,i2],[AP1,AP2],[LAN1,LAN2])
bodies = [x for x in particles if x.is_binary==False]
binaries = [x for x in particles if x.is_binary==True]

### Set integration terms ###
binaries[0].integration_method = 0 # orbit averaged
binaries[1].integration_method = 1 # direct integration (KS)

binaries[0].KS_use_perturbing_potential = True # toggle use KS perturbing potential (does not apply to this orbit, which is averaged)
binaries[1].KS_use_perturbing_potential = True # toggle use KS perturbing potential 

### Initialise the code ###
code = SecularMultiple()
code.add_particles(particles)

### Set the expansion order terms ###
code.include_quadrupole_order_terms = True # pairwise n=2
code.include_octupole_order_binary_pair_terms = True # pairwise n=3
code.include_octupole_order_binary_triplet_terms = True # triplet n=3
code.include_hexadecupole_order_binary_pair_terms = True # pairwise n=4
code.include_dotriacontupole_order_binary_pair_terms = True # pairwise n=5
code.include_double_averaging_corrections = False # whether or not to include averaging corrections (pairwise averaged; quadrupole order and test-particle limit)

### Run the code (dt and tend should be specified beforehand; note: the code determines its own internal timesteps dynamically; dt is the output timestep) ###
t = 0.0
while t<tend:
    code.evolve_model(t)
    t+=dt                    

    ### The following lines can be used to retrieve the orbital elements ###
    print("semimajor axes (AU) ", [x.a for x in binaries])
    print("eccentricities ", [x.e for x in binaries])
    print("inclinations (rad) ", [x.INCL for x in binaries])
    print("arguments of periapsis (rad) ", [AP for x in binaries])
    print("longitudes of the ascending node (rad) ", [x.INCL for x in binaries])
\end{lstlisting}

\subsection{Minimal use example}
\label{sect:use:ex}

Here, we briefly illustrate how to use the added features in \sm~in practice. A minimal \textsc{Python} example is given in Code Fragment~\ref{code:use}\footnote{We remark that the code examples given in Paper II were based on the \textsc{AMUSE} \citep{2013CoPhC.183..456P,2013A&A...557A..84P} version of \sm. The current updates apply to the standalone version of \sm, which, although very similar, has some differences in the interface. We recommend that users review the provided test and example scripts.}. The function \texttt{create\_nested\_multiple}, part of the included \texttt{Tools}, generates a set of particles representing a fully-nested hierarchical system (i.e., maximising the number of levels; in this case, for given $N$ bodies, the number of different levels is $N-2$). In the code fragment, a hierarchical triple is initialised. For future reference, particles representing bodies and binaries are separated out into the lists \texttt{bodies} and \texttt{binaries}, respectively. The integration method of each orbit is specified with the property \texttt{integration\_method}. The following options are implemented:

\begin{enumerate}
\item \texttt{integration\_method=0}: the orbit is averaged over (default value);
\item \texttt{integration\_method=1}: the orbit is integrated directly, using the KS element equations of motion;
\item \texttt{integration\_method=2}: the orbit is integrated directly, using the nonregularised approach (as described at the beginning of \S~\ref{sect:meth:dir}).
\end{enumerate}

For each binary, the bool \texttt{KS\_use\_perturbing\_potential} sets whether the perturbing potential or acceleration formulations are used (see \S~\ref{sect:meth:dir}). If \texttt{KS\_use\_perturbing\_potential=True}, the potential formulation is used; if \texttt{KS\_use\_perturbing\_potential=False}, the perturbing acceleration formulation is used.

Which expansion terms are included (for both averaged, and nonaveraged orbits) is specified with the code Boolean parameters 
\begin{itemize}
\item \texttt{code.include\_quadrupole\_order\_terms} (pairwise $n=2$);
\item \texttt{code.include\_octupole\_order\_binary\_pair\_terms}  (pairwise $n=3$);
\item \texttt{code.include\_octupole\_order\_binary\_triplet\_terms} (triplet $n=3$);
\item \texttt{code.include\_hexadecupole\_order\_binary\_pair\_terms} (pairwise $n=4$), and 
\item \texttt{code.include\_dotriacontupole\_order\_binary\_pair\_\newline terms} (pairwise $n=5$). 
\end{itemize}
The currently maximum supported expansion order for pairwise interactions is $n=5$, and $n=3$ for triplet interactions. By default, all these terms are enabled. 

The code parameter \texttt{code.include\_double\_averaging\_\newline corrections} determines whether or not orbit averaging corrections are included (see \S~\ref{sect:meth:cor}). By default, it is disabled. 

Integration of the system is achieved by running a time loop and using the code function \texttt{code.evolve\_model}. When a stopping condition is used (not included in the example in Code Fragment~\ref{code:use}), the stopping condition flag and time of stopping condition can be retrieved from \texttt{code.flag} and \texttt{code.model\_time}, respectively (refer to the included example and test scripts for examples on how to use stopping conditions in \sm). Orbital information can be retrieved from the code using the previously-defined \texttt{binaries} list.

\subsection{Determining which orbits to average over, and which to integrate directly}
\label{sect:use:choice}
The hybrid integration techniques and orbit-averaging corrections as presented above introduce potentially many more intricacies when integrating the long-term evolution of hierarchical systems using \sm. Here, we give some general recommendations for appropriate choices of which methods should be used in which situation. 

As a rule of thumb, averaging for a particular orbit breaks down when the timescale for the eccentricity and/or angular momentum vectors to change appreciably is comparable to, or shorter than the orbital timescale (e.g., \citealt{2014ApJ...781...45A}). The former timescale can be estimated as the LK timescale, which, for an orbit pair $(p,k)$, is given within an order of magnitude by (see, e.g., Paper I), 
\begin{align}
\label{eq:P_LK}
P_{\mathrm{LK},pk} \sim \frac{P_{\mathrm{orb},k}^2}{P_{\mathrm{orb},p}} \frac{M_k}{M_{k.\mathrm{CS}(p)}} \left (1-e_k^2 \right )^{3/2}.
\end{align}
Here $P_{\mathrm{orb},i}$ denotes the orbital period of orbit $i$. If $P_{\mathrm{LK},pk}$ for any pair in the system is comparable to or shorter than any orbital period in the system, then this is an indication that orbit averaging could break down. The longest orbital period is not necessarily an orbit associated with the pair $(p,k)$ used to evaluate $P_{\mathrm{LK},pk}$. For example, in 3+1 quadruple systems, the LK timescale associated with the innermost and intermediate orbits can be shorter than the outermost orbital period. Specifically, denoting the innermost, intermediate, and outer orbits with the labels 1, 2, and 3, respectively,
\begin{align}
\label{eq:a_quad_thres}
a_\mathrm{1} > a_\mathrm{2} \left ( \frac{a_\mathrm{2}}{a_\mathrm{3}}\right ) \left ( \frac{(m_1+m_2)(m_1+m_2+m_3+m_4)}{m_3^2} \right )^{1/3} \left(1-e_\mathrm{2}^2\right ),
\end{align}
where $m_1$ and $m_2$ are the component masses in the innermost binary, $m_3$ the mass of the intermediate body, and $m_4$ the mass of the outermost body. For example, the threshold $a_\mathrm{1}$ for the orbit-averaging breakdown is $a_\mathrm{1} \gtrsim 0.1 \, a_\mathrm{2}$ assuming $a_\mathrm{3}/a_\mathrm{2}=10$ and setting the other factors in \eq~(\ref{eq:a_quad_thres}) to unity. 

Generally, these timescales should be evaluated on a case-by-case basis, and they can be used to assess which orbits should be averaged over, and which should be integrated directly.

\section{Examples}
\label{sect:ex}
In this section, we present a number of examples in which the added features in \sm~can be beneficial: triples (\S~\ref{sect:ex:triple}), and quadruples (\S~\ref{sect:ex:quad}). The initial conditions for all the examples are listed in Table~\ref{table:values}.

\subsection{Triple systems}
\label{sect:ex:triple}
\subsubsection{Test system}
\label{sect:ex:triple:test}

\begin{figure}
\center
\includegraphics[scale = 0.44, trim = 0mm 10mm 0mm 10mm]{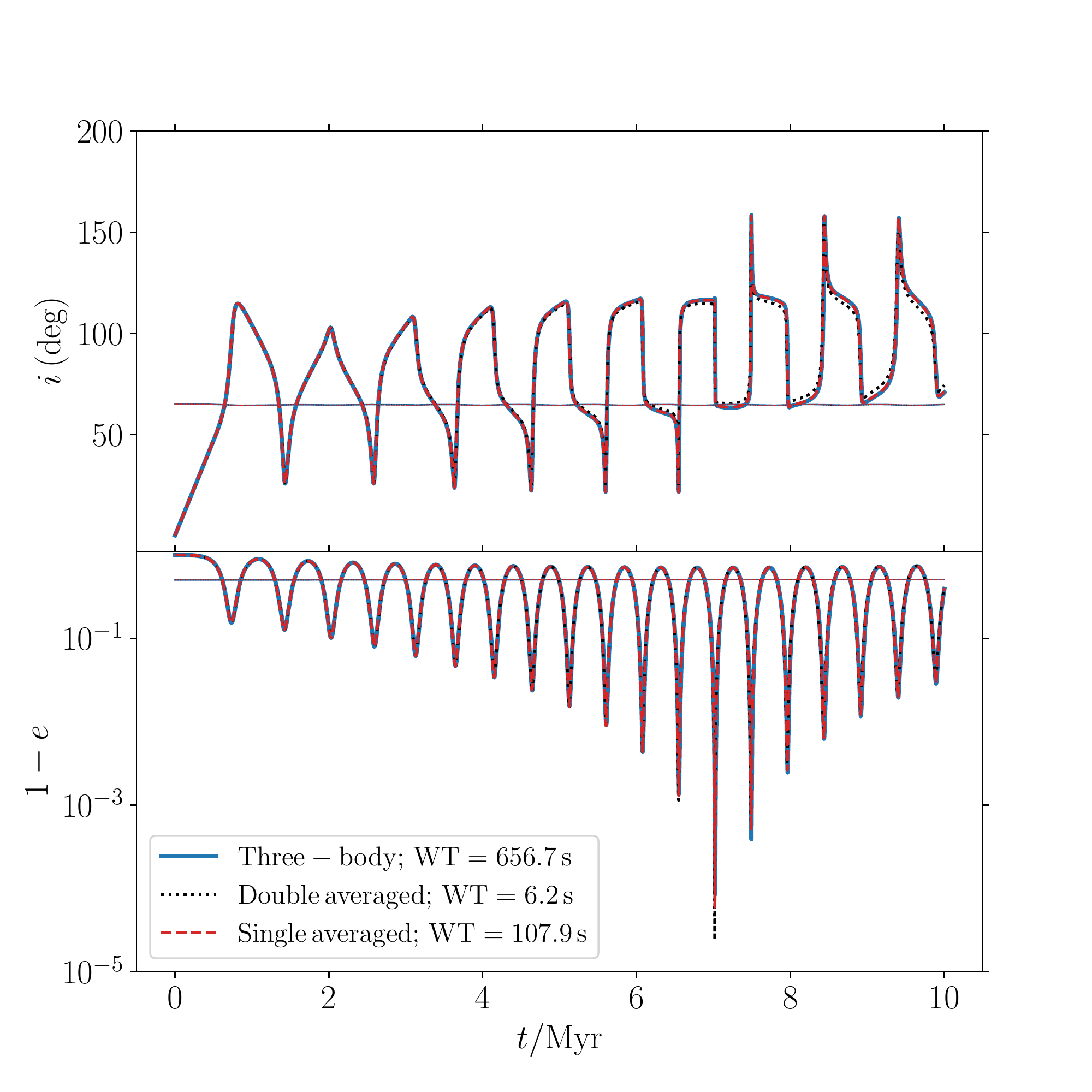}
\caption{Example evolution of a star-planet system orbited by a distant massive planet (see Table~\ref{table:values} for the initial conditions). Top panel: the individual inclinations of the inner and outer orbits (the outer orbit inclination remains nearly fixed at $65^\circ$). Bottom panel: the individual eccentricities (the outer orbit eccentricity remains nearly constant at $0.5$). Solid blue lines: direct integration with \textsc{REBOUND}. Black dotted lines: double averaged; red dashed lines: single averaged. The CPU wall time (`WT'; based on running on a single core on an Intel i9 9980HK) for each integration method is indicated in the legend. }
\label{fig:ex1}
\end{figure}

\F~\ref{fig:ex1} shows the secular evolution of a star-planet system orbited by a distant massive planet, with the initial conditions adopted from \citet{2013MNRAS.431.2155N}. We include three different integration methods: direct three-body integration using the \textsc{IAS15} integrator in \textsc{REBOUND} \citep{2012A&A...537A.128R,2015MNRAS.446.1424R}, shown with solid blue lines, double averaged (black dotted lines), and single averaged (red dashed lines). In this system, both orbital periods ($\simeq 15\,\yr$ and $\simeq 10^3\,\yr$ for the inner and outer orbits, respectively) are much shorter than the LK timescale (\eq~\ref{eq:P_LK}), which is $\simeq 1.1\,\Myr$. The double averaging approximation is therefore well justified, and this is reflected by the good agreement between the direct integration and double averaging, as well as single averaging methods. In the double averaging approximation, the maximum eccentricity near $t=7\,\Myr$ is slightly over-predicted, however. The single-averaging approximation yields a maximum eccentricity near this time which is closer to the direct three-body result. 

In terms of computational time, it is clear that direct integration is most costly, with a CPU wall time which is approximately six times longer compared to single averaging, and 100 times longer compared to double averaging. In this example, it is clear that the double averaging approximation offers a significant performance increase, while still being reasonably accurate. 

\begin{figure}
\center
\includegraphics[scale = 0.42, trim = 0mm 0mm 0mm 10mm]{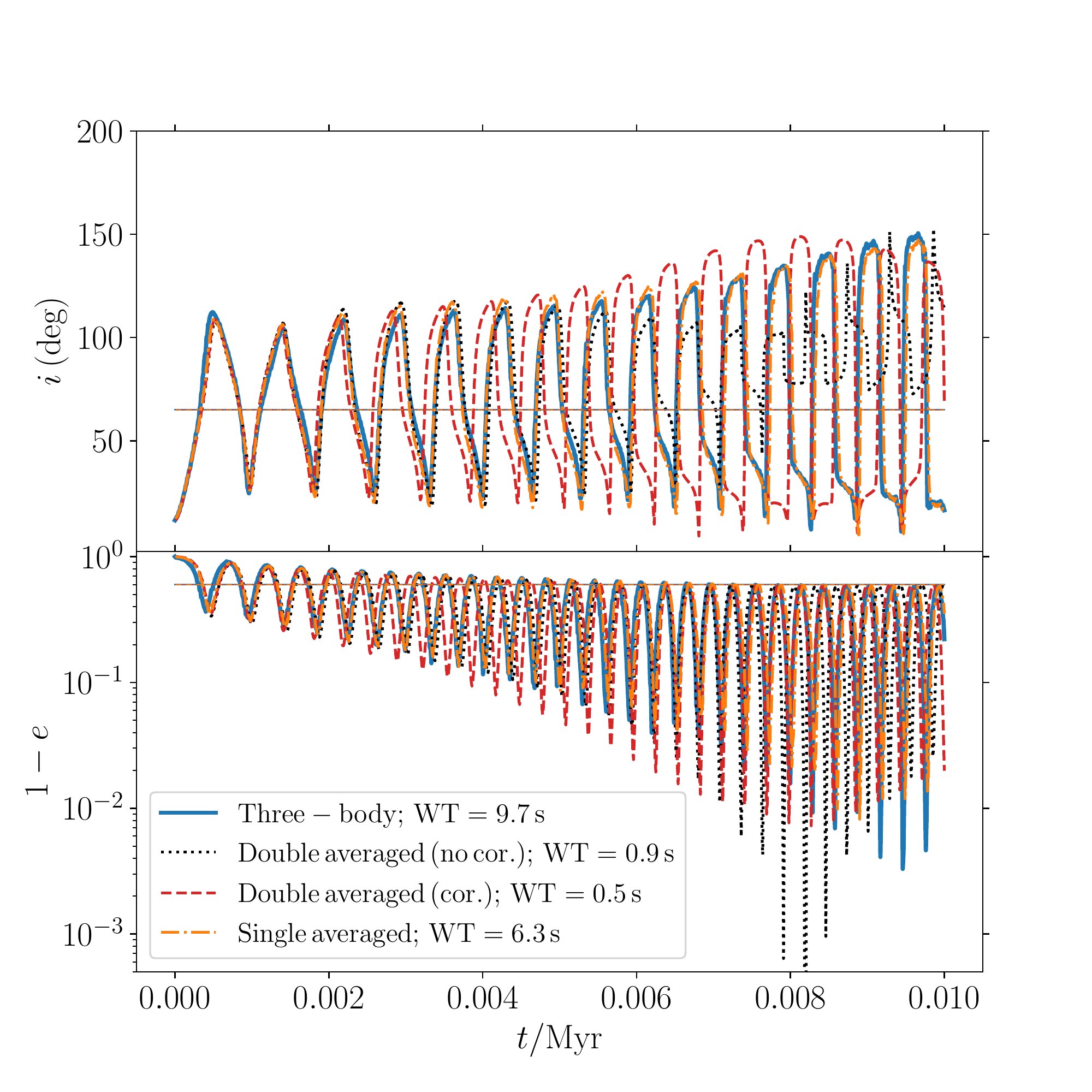}
\includegraphics[scale = 0.42, trim = 0mm 10mm 0mm 0mm]{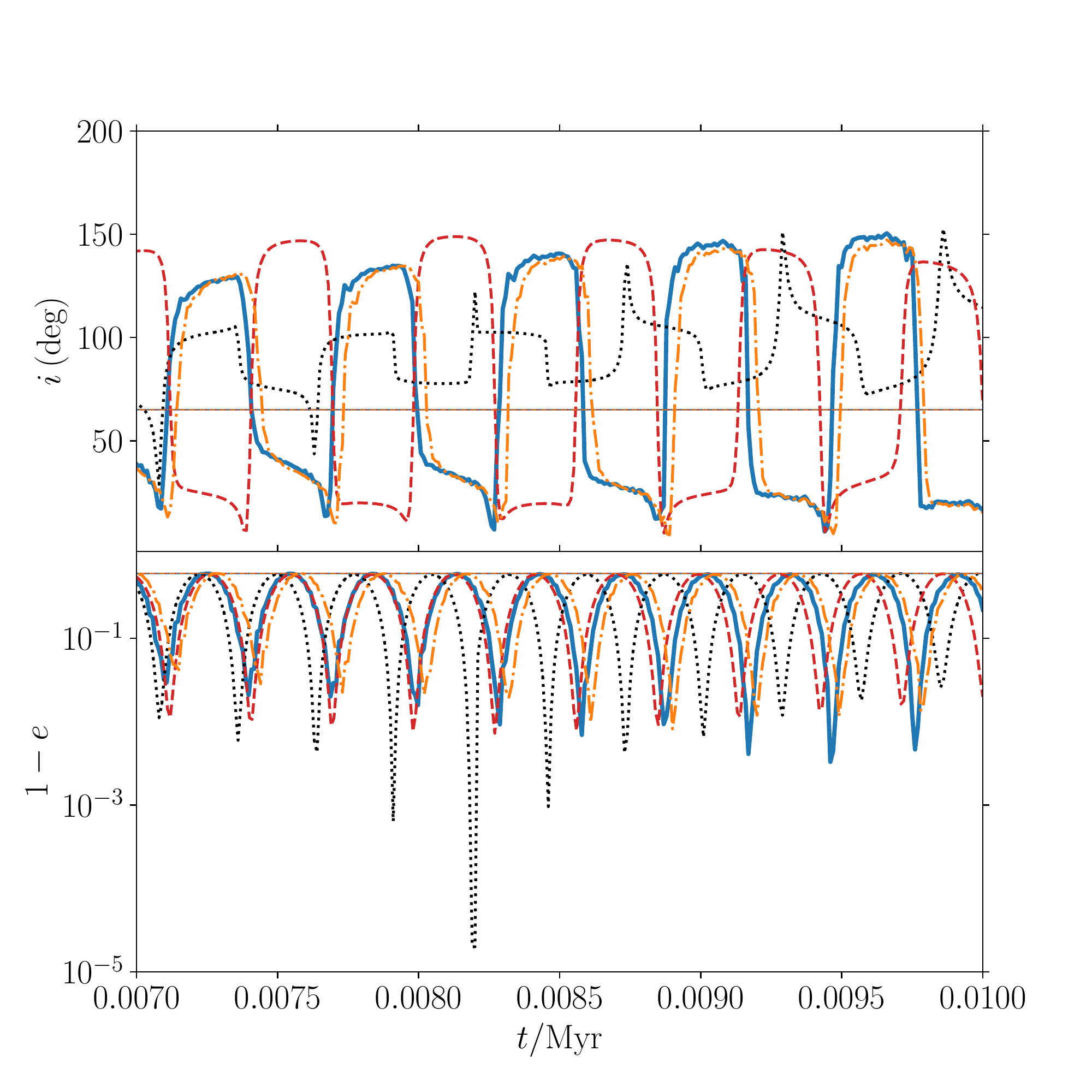}
\caption{Example evolution of a marginally hierarchical triple system (see Table~\ref{table:values} for the initial conditions). Top panel in the upper figure: the individual inclinations of the inner (thick lines) and outer (thin lines) orbits (the outer orbit inclination remains nearly fixed at $65^\circ$). Bottom panel in the upper figure: the individual eccentricities, with thick and thin lines corresponding to the inner and outer orbits, respectively (the outer orbit eccentricity remains nearly constant at $0.4$). The bottom figure shows the same data for the shorter timespan between 0.007 and 0.01 Myr. Solid blue lines: direct integration with \textsc{REBOUND}. Black dotted lines: double averaged without orbit-averaging correction terms; red dashed lines: double averaged including correction terms. Orange dot-dashed lines: single averaged. }
\label{fig:ex2}
\end{figure}

\subsubsection{Orbit-averaging corrections}
\label{sect:ex:triple:cor}

\F~\ref{fig:ex2} shows the example evolution of another hierarchical triple system. The orbital periods are $\simeq 1\,\yr$ and $\simeq22\,\yr$ for the inner and outer orbits, respectively; the LK timescale is $\simeq 770\,\yr$. With a semimajor axis ratio of 1/10 and an outer orbit eccentricity of $0.4$, this system is strongly interacting. With the double averaging approximation and without orbit-averaging corrections (black dotted lines), the predicted inner orbit eccentricity (thick lines) exceeds $1-10^{-3}$, which is inconsistent with the three-body integration (thick blue solid lines), for which the inner orbit eccentricity does not exceeds $1-10^{-2}$. This is no longer the case when orbit-averaging corrections are included (thick red dashed lines), illustrating the usefulness of the correction terms. With single averaging integration, the inner orbital eccentricity also does not exceed $1-10^{-2}$, although the CPU wall time is significantly longer and only $\sim 30\%$ faster compared to direct three-body integration.

\subsection{Quadruple systems}
\label{sect:ex:quad}
\subsubsection{3+1 system}
\label{sect:ex:quad:3p1}

\begin{figure}
\center
\includegraphics[scale = 0.42, trim = 0mm 10mm 0mm 10mm]{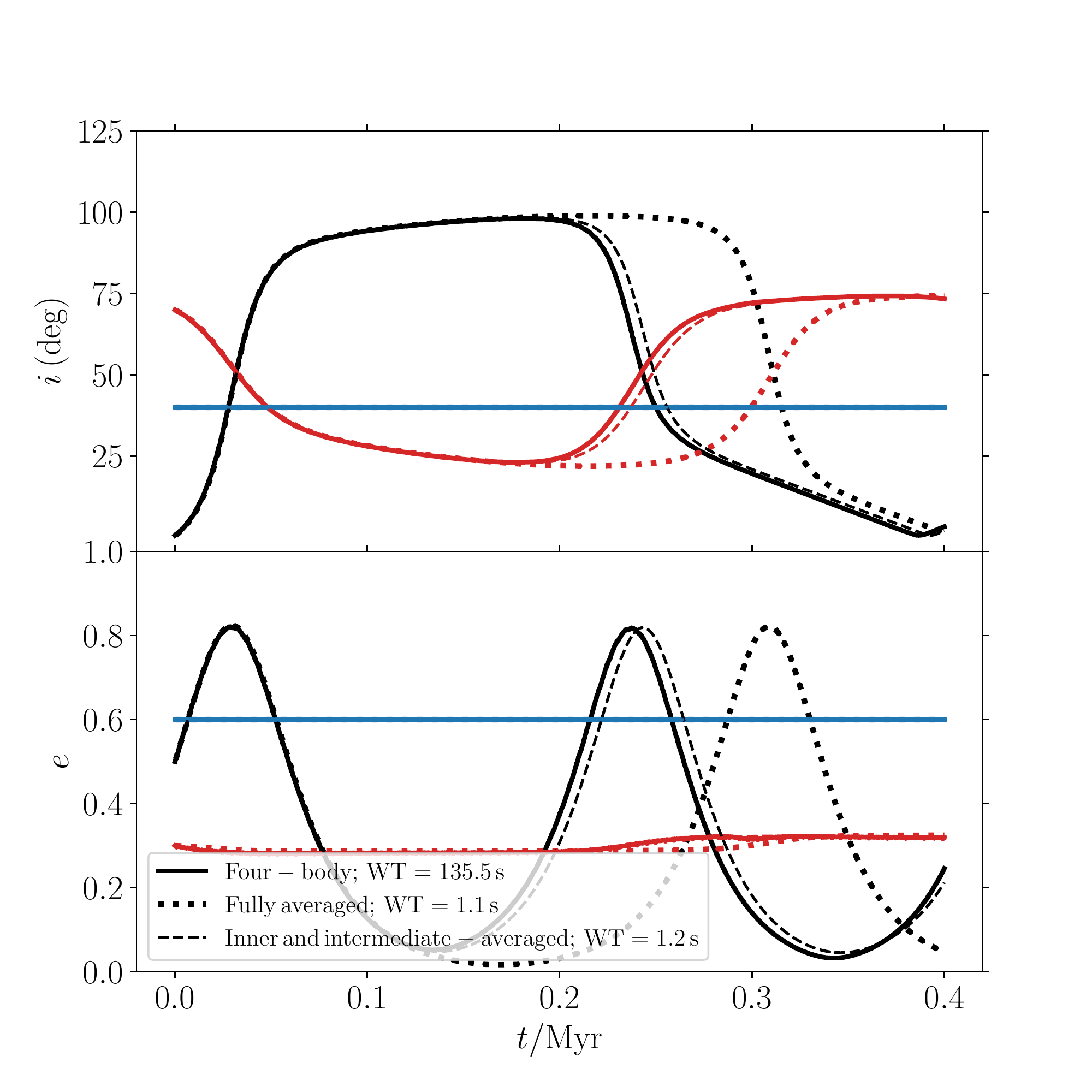}
\caption{ Short-term evolution of a 3+1 quadruple system (see Table~\ref{table:values} for the initial conditions). Top (bottom) panels show the individual inclinations (eccentricities); black lines correspond to the innermost orbit, red to the intermediate orbit, and blue to the outermost orbit. Solid lines: according to four-body integration (with \textsc{REBOUND}); dotted lines: fully averaged (over all three orbits); dashed lines: averaged over the inner and intermediate orbits, but integrating the outermost orbit directly. }
\label{fig:ex3}
\end{figure}

\F~\ref{fig:ex3} shows the short-term evolution of a 3+1 quadruple system. The orbital periods are $\simeq 29\,\yr$, $\simeq 877\,\yr$, and $\simeq 0.30\,\Myr$ for the inner, intermediate, and outermost orbits, respectively. The LK timescales associated with the inner-intermediate orbital pair is $t_\mathrm{LK,12} \simeq 0.30\,\Myr$; for the intermediate-outermost pair, it is $t_\mathrm{LK,23}\simeq 58 \,\Myr$. In this example, $t_\mathrm{LK,12}$ is very close to the outermost orbital period (in fact, the chosen value of $a_1$, $10\,\au$, is very nearly the same as the the critical value of $a_\mathrm{1}$ from \eq~\ref{eq:a_quad_thres}, which is $\simeq 10$). Therefore, it can be expected that the orbit-averaging approximation breaks down for the outermost orbit. 

This is indeed the case, since the fully-averaged integrations (dotted lines in \F~\ref{fig:ex3}) give a significantly different secular oscillation period compared to the direct four-body integration with \textsc{REBOUND} (solid lines). When averaging over the inner and intermediate orbits but directly integrating the outermost orbit (dashed lines), the agreement with the full direct integration becomes much better. Also, note the minimal computational impact of averaging over the outermost orbit versus integrating it directly, which can be explained by the very similar secular timescale and the outer orbital period. In contrast, full direct integration is more than 100 times slower compared to the hybrid approach.

\subsubsection{2+2 system}
\label{sect:ex:quad:2p2}

\begin{figure}
\center
\includegraphics[scale = 0.42, trim = 0mm 10mm 0mm 10mm]{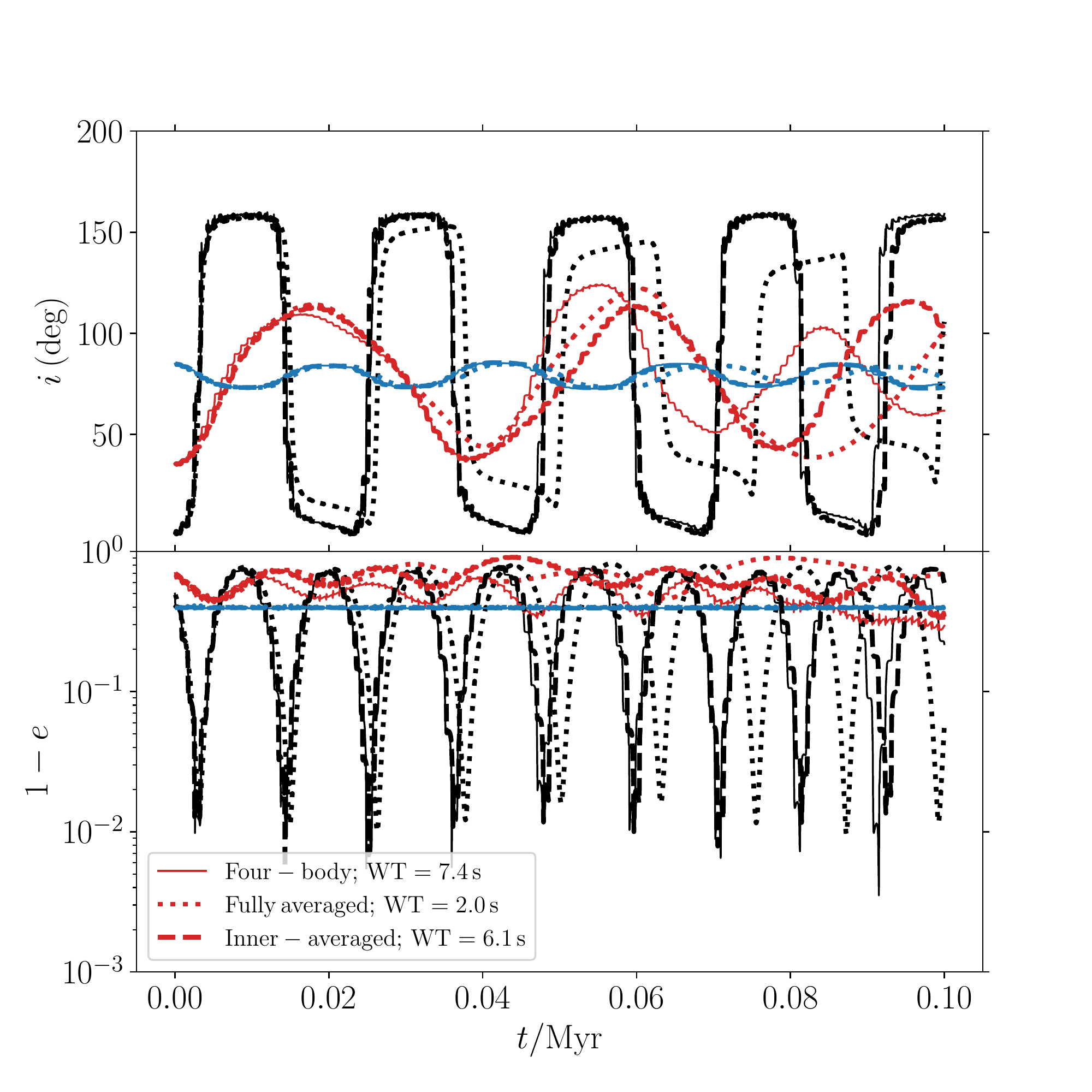}
\caption{ Short-term evolution of a 2+2 quadruple system (see Table~\ref{table:values} for the initial conditions). Top (bottom) panels show the individual inclinations (eccentricities); black lines correspond to the inner orbit labeled `1', red lines to the inner orbit labeled `2', and blue to the outer orbit. Solid lines: according to four-body integration (with \textsc{REBOUND}); dotted lines: fully averaged (over all three orbits); dashed lines: averaged over both inner orbits, but integrating the outer orbit directly. }
\label{fig:ex4}
\end{figure}

Lastly, we show in \F~\ref{fig:ex4} an example of a compact 2+2 quadruple system in which the period of the outer orbit (here labeled `3') is $\simeq 839\,\yr$, which is shorter than the LK timescales for both pairs, $t_\mathrm{LK,13} \simeq 28\,\kyr$, and $t_\mathrm{LK,23} \simeq 36\,\kyr$, but not by a large margin. Note that $t_\mathrm{LK,13}$ and $t_\mathrm{LK,23}$ are similar, implying that the system is secularly chaotic (e.g., \citealt{2017MNRAS.470.1657H}). 

The full averaging approximation (dotted lines) shows significant deviation from the direct four-body integration, in particular with respect to the period and magnitude of the eccentricity oscillations of inner orbit 1 (black lines). When averaging over both inner orbits but not the outer orbit (dashed lines), there is better agreement with the direct four-body integration, especially with respect to orbit 1. 

\begin{table*}
\begin{tabular}{lccccccccccccccccccc}
\toprule
& $m_1$ & $m_2$ & $m_3$ & $m_4$ & $a_1$ & $a_2$ & $a_3$ & $e_1$ & $e_2$ & $e_3$ & $i_1$ & $i_2$ & $i_3$ & $\omega_1$ & $\omega_2$ & $\omega_3$ & $\Omega_1$ & $\Omega_2$ & $\Omega_3$ \\
\midrule
\F~\ref{fig:ex1} & 1 & $10^{-3}$ & $40 \times 10^{-3}$ & --- & 6 & 100 & --- & 0 & 0.5 & --- & 0 & 65 & --- & 45 & 0 & --- & 0 & 0 & --- \\
\F~\ref{fig:ex2} & 1 & $10^{-6}$ & $1$ & --- & 1 & 10 & --- & 0 & 0.4 & --- & 11.5 & 65 & --- & 0 & 0 & --- & 0 & 0 & --- \\
\F~\ref{fig:ex3} & 1 & 0.2 & 0.1 & 10 & 10 & 100 & $10^4$ & 0.5 & 0.3 & 0.6 & 0.6 & 70 & 40 & 45 & 0.01 & 0.01 & 0.6 & 0.6 & 0.6 \\
\F~\ref{fig:ex4} & 1 & 0.8 & 0.1 & 2.0 & 10 & 10 & $140$ & 0.5 & 0.3 & 0.6 & 0.6 & 35 & 85 & 45 & 40 & 140 & 0.6 & 0.6 & 0.6 \\
\bottomrule
\end{tabular}
\caption{Values of parameters used in the examples. Masses $m_i$ are in Solar units, semimajor axes $a_i$ in units of $\au$; orbital angles (inclinations $i_i$, arguments of periapsis $\omega_i$, and longitudes of the ascending node $\Omega_i$) are measured in degrees (our reference frame is the $x,y$-plane, and the reference direction is the $x$-direction). Quantities that do not apply are indicated with `---'.}
\label{table:values}
\end{table*}

\section{Discussion}
\label{sect:discussion}
\subsection{Fully nonaveraged integration in relation to direct $N$-body integration}
\label{sect:dir}
The hybrid integration techniques in \sm~introduced here also allow for direct integration of {\it all} orbits in the system. We remark that, in our testing, this approach tends to be significantly slower than integrating the system with a direct $N$-body code using special integration techniques such as algorithmic regularisation (e.g., \citealt{2006MNRAS.372..219M,2008AJ....135.2398M,2020MNRAS.492.4131R}), possibly combined with the slowdown method (e.g., \citealt{1996CeMDA..64..197M,2020MNRAS.tmp..452W}). This may appear surprising, but we note that in \sm, direct integration is based on an expansion of the Hamiltonian (see \S~\ref{sect:meth:ham}). In particular when high expansion orders are included, this implies that the acceleration terms are computationally relatively expensive to evaluate. In other words, the Hamiltonian expansion is advantageous when averaging over at least one orbit, but is not suited when directly integrating all orbits. We therefore generally recommend using direct $N$-body integration methods when the intention is to integrate all orbits directly. 

\subsection{Orbit-averaging corrections}
\label{sect:discussion:cor}
As described in \S~\ref{sect:meth:cor}, the current implementation of the orbit-averaging correction terms in \sm~is based on the expressions of \citet{2016MNRAS.458.3060L}, which apply to hierarchical triples in the test-particle limit, and to the quadrupole expansion order. We generally recommend only applying them in systems in which those restrictions apply. For this reason, the correction terms are disabled by default in the code (see also \S~\ref{sect:use}). We defer a more self-consistent treatment of orbit-averaging correction terms that applies to more general systems to future work. 

\subsection{Short-range forces}
\label{sect:discussion:pn}
The current implementation of direct integration in \sm~does not include non-Newtonian point mass terms such as post-Newtonian terms or terms associated with tidal evolution, both of which are associated with short-range forces. Such terms are included for averaged orbits (see also Paper I), however. Since \sm~is generally not intended to be used with direct integration in the inner(most) orbits (cf. \S~\ref{sect:dir}), in the typical use case, integrated orbits are wider orbits for which short-range forces are less important (exceptions exist in high mass-ratio systems, such as binaries orbiting a supermassive black hole). For simplicity and performance reasons, we therefore currently do not include short-range forces for orbits that are integrated directly.

\section{Conclusions}
\label{sect:conclusions}
We presented an update to the \sm~code, which integrates the long-term dynamical evolution of multiple systems with any number of bodies and hierarchical structure, provided that the system is composed of nested binaries. Whereas previously we averaged the Hamiltonian over all orbits, we now also implemented hybrid integration methods. We also implemented orbit averaging corrections. Our main conclusions are listed below. 

\medskip \noindent 1. In the updated code, the user can specify for each orbit if the dynamical evolution should be modelled by averaging over it, or by direct integration (i.e., resolving its orbital motion). We derived the Hamiltonian and equations of motion for this hybrid orbit integration scheme for pairwise integrations to any expansion order $n$, and for triplet interactions (involving three orbits simultaneously) at the octupole order ($n=3$). In the code, pairwise interactions are included up to and including fifth order, and for triplet interactions to octupole order. 

\medskip \noindent 2. Hybrid integration is useful in situations when the orbit averaging approximation breaks down in some orbits, but is still valid in others. By effectively combining direct integration and orbit averaging the long-term evolution can be accurately modelled, but with significantly lower computational cost compared to existing direct $N$-body integration codes. To evaluate which orbits should be averaged over and which should be integrated directly, the secular evolution timescales should be compared to the orbital periods. For the averaging approximation to be valid in a particular orbit, its orbital timescale should be much shorter than the secular evolution timescale of other orbits in the system.

\medskip \noindent 3. We also incorporated analytical orbit-averaging corrections for pairwise interactions to quadrupole order, in the test-particle approximation (from \citealt{2016MNRAS.458.3060L}). These terms, although having limitations, can be used to model the secular evolution more accurately with no significant performance loss. 

\medskip \noindent 4. We discussed a number of examples (triples and quadruples) in which the added features presented here can be beneficial.

Our updated code, which is written in \textsc{C++} supplemented by a user-friendly interface in \textsc{Python}, is freely available (see the link provided at the end of \S~\ref{sect:introduction}).

\section*{Acknowledgements}
I thank Javier Roa for his expert advice on the application of the KS regularised element equations of motion, and the anonymous referee for useful suggestions that helped to improve the manuscript. Simulations in this paper made use of the \textsc{REBOUND} code which is freely available at \href{http://github.com/hannorein/rebound}{http://github.com/hannorein/rebound}. 

\bibliographystyle{mnras}
\bibliography{literature}

\begin{thebibliography}{}
\makeatletter
\relax
\def\mn@urlcharsother{\let\do\@makeother \do\$\do\&\do\#\do\^\do\_\do\%\do\~}
\def\mn@doi{\begingroup\mn@urlcharsother \@ifnextchar [ {\mn@doi@}
  {\mn@doi@[]}}
\def\mn@doi@[#1]#2{\def\@tempa{#1}\ifx\@tempa\@empty \href
  {http://dx.doi.org/#2} {doi:#2}\else \href {http://dx.doi.org/#2} {#1}\fi
  \endgroup}
\def\mn@eprint#1#2{\mn@eprint@#1:#2::\@nil}
\def\mn@eprint@arXiv#1{\href {http://arxiv.org/abs/#1} {{\tt arXiv:#1}}}
\def\mn@eprint@dblp#1{\href {http://dblp.uni-trier.de/rec/bibtex/#1.xml}
  {dblp:#1}}
\def\mn@eprint@#1:#2:#3:#4\@nil{\def\@tempa {#1}\def\@tempb {#2}\def\@tempc
  {#3}\ifx \@tempc \@empty \let \@tempc \@tempb \let \@tempb \@tempa \fi \ifx
  \@tempb \@empty \def\@tempb {arXiv}\fi \@ifundefined
  {mn@eprint@\@tempb}{\@tempb:\@tempc}{\expandafter \expandafter \csname
  mn@eprint@\@tempb\endcsname \expandafter{\@tempc}}}

\bibitem[\protect\citeauthoryear{{Allan} \& {Cook}}{{Allan} \&
  {Cook}}{1964}]{1964RSPSA.280...97A}
{Allan} R.~R.,  {Cook} G.~E.,  1964, \mn@doi [Royal Society of London
  Proceedings Series A] {10.1098/rspa.1964.0133}, \href
  {http://cdsads.u-strasbg.fr/abs/1964RSPSA.280...97A} {280, 97}

\bibitem[\protect\citeauthoryear{{Allan} \& {Ward}}{{Allan} \&
  {Ward}}{1963}]{1963PCPS...59..669A}
{Allan} R.~R.,  {Ward} G.~N.,  1963, \mn@doi [Cambridge Philosophical Society
  Proceedings] {10.1017/S0305004100037336}, \href
  {http://cdsads.u-strasbg.fr/abs/1963PCPS...59..669A} {59, 669}

\bibitem[\protect\citeauthoryear{{Anderson}, {Storch}  \& {Lai}}{{Anderson}
  et~al.}{2016}]{2016MNRAS.456.3671A}
{Anderson} K.~R.,  {Storch} N.~I.,   {Lai} D.,  2016, \mn@doi [\mnras]
  {10.1093/mnras/stv2906}, \href
  {http://cdsads.u-strasbg.fr/abs/2016MNRAS.456.3671A} {456, 3671}

\bibitem[\protect\citeauthoryear{{Antognini}}{{Antognini}}{2016}]{2016PhDT........82A}
{Antognini} J.~M.,  2016, PhD thesis, The Ohio State University

\bibitem[\protect\citeauthoryear{{Antonini} \& {Perets}}{{Antonini} \&
  {Perets}}{2012}]{2012ApJ...757...27A}
{Antonini} F.,  {Perets} H.~B.,  2012, \mn@doi [\apj]
  {10.1088/0004-637X/757/1/27}, \href
  {http://adsabs.harvard.edu/abs/2012ApJ...757...27A} {757, 27}

\bibitem[\protect\citeauthoryear{{Antonini}, {Murray}  \& {Mikkola}}{{Antonini}
  et~al.}{2014}]{2014ApJ...781...45A}
{Antonini} F.,  {Murray} N.,   {Mikkola} S.,  2014, \mn@doi [\apj]
  {10.1088/0004-637X/781/1/45}, \href
  {http://adsabs.harvard.edu/abs/2014ApJ...781...45A} {781, 45}

\bibitem[\protect\citeauthoryear{{Antonini}, {Chatterjee}, {Rodriguez},
  {Morscher}, {Pattabiraman}, {Kalogera}  \& {Rasio}}{{Antonini}
  et~al.}{2016}]{2016ApJ...816...65A}
{Antonini} F.,  {Chatterjee} S.,  {Rodriguez} C.~L.,  {Morscher} M.,
  {Pattabiraman} B.,  {Kalogera} V.,   {Rasio} F.~A.,  2016, \mn@doi [\apj]
  {10.3847/0004-637X/816/2/65}, \href
  {http://adsabs.harvard.edu/abs/2016ApJ...816...65A} {816, 65}

\bibitem[\protect\citeauthoryear{{Antonini}, {Toonen}  \& {Hamers}}{{Antonini}
  et~al.}{2017}]{2017ApJ...841...77A}
{Antonini} F.,  {Toonen} S.,   {Hamers} A.~S.,  2017, \mn@doi [\apj]
  {10.3847/1538-4357/aa6f5e}, \href
  {http://adsabs.harvard.edu/abs/2017ApJ...841...77A} {841, 77}

\bibitem[\protect\citeauthoryear{{Blaes}, {Lee}  \& {Socrates}}{{Blaes}
  et~al.}{2002}]{2002ApJ...578..775B}
{Blaes} O.,  {Lee} M.~H.,   {Socrates} A.,  2002, \mn@doi [\apj]
  {10.1086/342655}, \href {http://adsabs.harvard.edu/abs/2002ApJ...578..775B}
  {578, 775}

\bibitem[\protect\citeauthoryear{{Breiter} \& {Ratajczak}}{{Breiter} \&
  {Ratajczak}}{2005}]{2005MNRAS.364.1222B}
{Breiter} S.,  {Ratajczak} R.,  2005, \mn@doi [\mnras]
  {10.1111/j.1365-2966.2005.09658.x}, \href
  {http://cdsads.u-strasbg.fr/abs/2005MNRAS.364.1222B} {364, 1222}

\bibitem[\protect\citeauthoryear{{Eggleton} \& {Kiseleva-Eggleton}}{{Eggleton}
  \& {Kiseleva-Eggleton}}{2001}]{2001ApJ...562.1012E}
{Eggleton} P.~P.,  {Kiseleva-Eggleton} L.,  2001, \mn@doi [\apj]
  {10.1086/323843}, \href {http://adsabs.harvard.edu/abs/2001ApJ...562.1012E}
  {562, 1012}

\bibitem[\protect\citeauthoryear{{Eggleton} \& {Kisseleva-Eggleton}}{{Eggleton}
  \& {Kisseleva-Eggleton}}{2006}]{2006Ap&SS.304...75E}
{Eggleton} P.~P.,  {Kisseleva-Eggleton} L.,  2006, \mn@doi [\apss]
  {10.1007/s10509-006-9078-z}, \href
  {http://adsabs.harvard.edu/abs/2006Ap%26SS.304...75E} {304, 75}

\bibitem[\protect\citeauthoryear{{Fabrycky} \& {Tremaine}}{{Fabrycky} \&
  {Tremaine}}{2007}]{2007ApJ...669.1298F}
{Fabrycky} D.,  {Tremaine} S.,  2007, \mn@doi [\apj] {10.1086/521702}, \href
  {http://adsabs.harvard.edu/abs/2007ApJ...669.1298F} {669, 1298}

\bibitem[\protect\citeauthoryear{{Fang}, {Thompson}  \& {Hirata}}{{Fang}
  et~al.}{2018}]{2018MNRAS.476.4234F}
{Fang} X.,  {Thompson} T.~A.,   {Hirata} C.~M.,  2018, \mn@doi [\mnras]
  {10.1093/mnras/sty472}, \href
  {https://ui.adsabs.harvard.edu/abs/2018MNRAS.476.4234F} {476, 4234}

\bibitem[\protect\citeauthoryear{{Ford}, {Kozinsky}  \& {Rasio}}{{Ford}
  et~al.}{2000}]{2000ApJ...535..385F}
{Ford} E.~B.,  {Kozinsky} B.,   {Rasio} F.~A.,  2000, \mn@doi [\apj]
  {10.1086/308815}, \href {http://adsabs.harvard.edu/abs/2000ApJ...535..385F}
  {535, 385}

\bibitem[\protect\citeauthoryear{{Fragione} \& {Antonini}}{{Fragione} \&
  {Antonini}}{2019}]{2019MNRAS.488..728F}
{Fragione} G.,  {Antonini} F.,  2019, \mn@doi [\mnras] {10.1093/mnras/stz1723},
  \href {https://ui.adsabs.harvard.edu/abs/2019MNRAS.488..728F} {488, 728}

\bibitem[\protect\citeauthoryear{{Fragione} \& {Kocsis}}{{Fragione} \&
  {Kocsis}}{2019}]{2019MNRAS.486.4781F}
{Fragione} G.,  {Kocsis} B.,  2019, \mn@doi [\mnras] {10.1093/mnras/stz1175},
  \href {https://ui.adsabs.harvard.edu/abs/2019MNRAS.486.4781F} {486, 4781}

\bibitem[\protect\citeauthoryear{{Fragione} \& {Loeb}}{{Fragione} \&
  {Loeb}}{2019}]{2019MNRAS.486.4443F}
{Fragione} G.,  {Loeb} A.,  2019, \mn@doi [\mnras] {10.1093/mnras/stz1131},
  \href {https://ui.adsabs.harvard.edu/abs/2019MNRAS.486.4443F} {486, 4443}

\bibitem[\protect\citeauthoryear{{Franchini}, {Martin}  \& {Lubow}}{{Franchini}
  et~al.}{2019}]{2019MNRAS.485..315F}
{Franchini} A.,  {Martin} R.~G.,   {Lubow} S.~H.,  2019, \mn@doi [\mnras]
  {10.1093/mnras/stz424}, \href
  {https://ui.adsabs.harvard.edu/abs/2019MNRAS.485..315F} {485, 315}

\bibitem[\protect\citeauthoryear{{Fu}, {Lubow}  \& {Martin}}{{Fu}
  et~al.}{2015}]{2015ApJ...813..105F}
{Fu} W.,  {Lubow} S.~H.,   {Martin} R.~G.,  2015, \mn@doi [\apj]
  {10.1088/0004-637X/813/2/105}, \href
  {https://ui.adsabs.harvard.edu/abs/2015ApJ...813..105F} {813, 105}

\bibitem[\protect\citeauthoryear{{Grishin}, {Lai}  \& {Perets}}{{Grishin}
  et~al.}{2018a}]{2018MNRAS.474.3547G}
{Grishin} E.,  {Lai} D.,   {Perets} H.~B.,  2018a, \mn@doi [\mnras]
  {10.1093/mnras/stx3005}, \href
  {http://adsabs.harvard.edu/abs/2018MNRAS.474.3547G} {474, 3547}

\bibitem[\protect\citeauthoryear{{Grishin}, {Perets}  \& {Fragione}}{{Grishin}
  et~al.}{2018b}]{2018MNRAS.481.4907G}
{Grishin} E.,  {Perets} H.~B.,   {Fragione} G.,  2018b, \mn@doi [\mnras]
  {10.1093/mnras/sty2477}, \href
  {http://adsabs.harvard.edu/abs/2018MNRAS.481.4907G} {481, 4907}

\bibitem[\protect\citeauthoryear{{Guenther}, {Hartmann}, {Esposito}, {Hatzes},
  {Cusano}  \& {Gandolfi}}{{Guenther} et~al.}{2009}]{2009A&A...507.1659G}
{Guenther} E.~W.,  {Hartmann} M.,  {Esposito} M.,  {Hatzes} A.~P.,  {Cusano}
  F.,   {Gandolfi} D.,  2009, \mn@doi [\aap] {10.1051/0004-6361/200912112},
  \href {http://adsabs.harvard.edu/abs/2009A%26A...507.1659G} {507, 1659}

\bibitem[\protect\citeauthoryear{{Hamers}}{{Hamers}}{2018}]{2018MNRAS.476.4139H}
{Hamers} A.~S.,  2018, \mn@doi [\mnras] {10.1093/mnras/sty428}, \href
  {https://ui.adsabs.harvard.edu/\#abs/2018MNRAS.476.4139H} {476, 4139}

\bibitem[\protect\citeauthoryear{{Hamers}}{{Hamers}}{2020}]{2020arXiv200208746H}
{Hamers} A.~S.,  2020, arXiv e-prints, \href
  {https://ui.adsabs.harvard.edu/abs/2020arXiv200208746H} {p. arXiv:2002.08746}

\bibitem[\protect\citeauthoryear{{Hamers} \& {Lai}}{{Hamers} \&
  {Lai}}{2017}]{2017MNRAS.470.1657H}
{Hamers} A.~S.,  {Lai} D.,  2017, \mn@doi [\mnras] {10.1093/mnras/stx1319},
  \href {http://adsabs.harvard.edu/abs/2017MNRAS.470.1657H} {470, 1657}

\bibitem[\protect\citeauthoryear{{Hamers} \& {Portegies Zwart}}{{Hamers} \&
  {Portegies Zwart}}{2016}]{2016MNRAS.459.2827H}
{Hamers} A.~S.,  {Portegies Zwart} S.~F.,  2016, \mn@doi [\mnras]
  {10.1093/mnras/stw784}, \href
  {http://adsabs.harvard.edu/abs/2016MNRAS.459.2827H} {459, 2827}

\bibitem[\protect\citeauthoryear{{Hamers}, {Pols}, {Claeys}  \&
  {Nelemans}}{{Hamers} et~al.}{2013}]{2013MNRAS.430.2262H}
{Hamers} A.~S.,  {Pols} O.~R.,  {Claeys} J.~S.~W.,   {Nelemans} G.,  2013,
  \mn@doi [\mnras] {10.1093/mnras/stt046}, \href
  {https://ui.adsabs.harvard.edu/abs/2013MNRAS.430.2262H} {430, 2262}

\bibitem[\protect\citeauthoryear{{Hamers}, {Perets}, {Antonini}  \& {Portegies
  Zwart}}{{Hamers} et~al.}{2015}]{2015MNRAS.449.4221H}
{Hamers} A.~S.,  {Perets} H.~B.,  {Antonini} F.,   {Portegies Zwart} S.~F.,
  2015, \mn@doi [\mnras] {10.1093/mnras/stv452}, \href
  {https://ui.adsabs.harvard.edu/abs/2015MNRAS.449.4221H} {449, 4221}

\bibitem[\protect\citeauthoryear{{Harrington}}{{Harrington}}{1968}]{1968AJ.....73..190H}
{Harrington} R.~S.,  1968, \mn@doi [\aj] {10.1086/110614}, \href
  {http://adsabs.harvard.edu/abs/1968AJ.....73..190H} {73, 190}

\bibitem[\protect\citeauthoryear{{Hoang}, {Naoz}, {Kocsis}, {Rasio}  \&
  {Dosopoulou}}{{Hoang} et~al.}{2018}]{2018ApJ...856..140H}
{Hoang} B.-M.,  {Naoz} S.,  {Kocsis} B.,  {Rasio} F.~A.,   {Dosopoulou} F.,
  2018, \mn@doi [\apj] {10.3847/1538-4357/aaafce}, \href
  {http://adsabs.harvard.edu/abs/2018ApJ...856..140H} {856, 140}

\bibitem[\protect\citeauthoryear{{Ito} \& {Ohtsuka}}{{Ito} \&
  {Ohtsuka}}{2019}]{2019MEEP....7....1I}
{Ito} T.,  {Ohtsuka} K.,  2019, \mn@doi [Monographs on Environment, Earth and
  Planets] {10.5047/meep.2019.00701.0001}, \href
  {https://ui.adsabs.harvard.edu/abs/2019MEEP....7....1I} {7, 1}

\bibitem[\protect\citeauthoryear{{Kiseleva}, {Eggleton}  \&
  {Mikkola}}{{Kiseleva} et~al.}{1998}]{1998MNRAS.300..292K}
{Kiseleva} L.~G.,  {Eggleton} P.~P.,   {Mikkola} S.,  1998, \mn@doi [\mnras]
  {10.1046/j.1365-8711.1998.01903.x}, \href
  {http://cdsads.u-strasbg.fr/abs/1998MNRAS.300..292K} {300, 292}

\bibitem[\protect\citeauthoryear{{Kozai}}{{Kozai}}{1962}]{1962AJ.....67..591K}
{Kozai} Y.,  1962, \mn@doi [\aj] {10.1086/108790}, \href
  {http://adsabs.harvard.edu/abs/1962AJ.....67..591K} {67, 591}

\bibitem[\protect\citeauthoryear{{Lei}}{{Lei}}{2019}]{2019MNRAS.490.4756L}
{Lei} H.,  2019, \mn@doi [\mnras] {10.1093/mnras/stz2917}, \href
  {https://ui.adsabs.harvard.edu/abs/2019MNRAS.490.4756L} {490, 4756}

\bibitem[\protect\citeauthoryear{{Lei}, {Circi}  \& {Ortore}}{{Lei}
  et~al.}{2018}]{2018MNRAS.481.4602L}
{Lei} H.,  {Circi} C.,   {Ortore} E.,  2018, \mn@doi [\mnras]
  {10.1093/mnras/sty2619}, \href
  {http://adsabs.harvard.edu/abs/2018MNRAS.481.4602L} {481, 4602}

\bibitem[\protect\citeauthoryear{{Lidov}}{{Lidov}}{1962}]{1962P&SS....9..719L}
{Lidov} M.~L.,  1962, \mn@doi [\planss] {10.1016/0032-0633(62)90129-0}, \href
  {http://adsabs.harvard.edu/abs/1962P%26SS....9..719L} {9, 719}

\bibitem[\protect\citeauthoryear{{Liu} \& {Lai}}{{Liu} \&
  {Lai}}{2017}]{2017ApJ...846L..11L}
{Liu} B.,  {Lai} D.,  2017, \mn@doi [\apjl] {10.3847/2041-8213/aa8727}, \href
  {http://adsabs.harvard.edu/abs/2017ApJ...846L..11L} {846, L11}

\bibitem[\protect\citeauthoryear{{Liu} \& {Lai}}{{Liu} \&
  {Lai}}{2018}]{2018ApJ...863...68L}
{Liu} B.,  {Lai} D.,  2018, \mn@doi [\apj] {10.3847/1538-4357/aad09f}, \href
  {https://ui.adsabs.harvard.edu/abs/2018ApJ...863...68L} {863, 68}

\bibitem[\protect\citeauthoryear{{Liu} \& {Lai}}{{Liu} \&
  {Lai}}{2019}]{2019MNRAS.483.4060L}
{Liu} B.,  {Lai} D.,  2019, \mn@doi [\mnras] {10.1093/mnras/sty3432}, \href
  {https://ui.adsabs.harvard.edu/abs/2019MNRAS.483.4060L} {483, 4060}

\bibitem[\protect\citeauthoryear{{Lubow} \& {Ogilvie}}{{Lubow} \&
  {Ogilvie}}{2017}]{2017MNRAS.469.4292L}
{Lubow} S.~H.,  {Ogilvie} G.~I.,  2017, \mn@doi [\mnras]
  {10.1093/mnras/stx990}, \href
  {https://ui.adsabs.harvard.edu/abs/2017MNRAS.469.4292L} {469, 4292}

\bibitem[\protect\citeauthoryear{{Luo}, {Katz}  \& {Dong}}{{Luo}
  et~al.}{2016}]{2016MNRAS.458.3060L}
{Luo} L.,  {Katz} B.,   {Dong} S.,  2016, \mn@doi [\mnras]
  {10.1093/mnras/stw475}, \href
  {http://cdsads.u-strasbg.fr/abs/2016MNRAS.458.3060L} {458, 3060}

\bibitem[\protect\citeauthoryear{{Martin} \& {Franchini}}{{Martin} \&
  {Franchini}}{2019}]{2019MNRAS.489.1797M}
{Martin} R.~G.,  {Franchini} A.,  2019, \mn@doi [\mnras]
  {10.1093/mnras/stz2250}, \href
  {https://ui.adsabs.harvard.edu/abs/2019MNRAS.489.1797M} {489, 1797}

\bibitem[\protect\citeauthoryear{{Martin}, {Nixon}, {Lubow}, {Armitage},
  {Price}, {Do{\u g}an}  \& {King}}{{Martin}
  et~al.}{2014}]{2014ApJ...792L..33M}
{Martin} R.~G.,  {Nixon} C.,  {Lubow} S.~H.,  {Armitage} P.~J.,  {Price} D.~J.,
   {Do{\u g}an} S.,   {King} A.,  2014, \mn@doi [\apjl]
  {10.1088/2041-8205/792/2/L33}, \href
  {http://adsabs.harvard.edu/abs/2014ApJ...792L..33M} {792, L33}

\bibitem[\protect\citeauthoryear{{Mazeh} \& {Shaham}}{{Mazeh} \&
  {Shaham}}{1979}]{1979A&A....77..145M}
{Mazeh} T.,  {Shaham} J.,  1979, \aap, \href
  {http://cdsads.u-strasbg.fr/abs/1979A%26A....77..145M} {77, 145}

\bibitem[\protect\citeauthoryear{{Mikkola} \& {Aarseth}}{{Mikkola} \&
  {Aarseth}}{1996}]{1996CeMDA..64..197M}
{Mikkola} S.,  {Aarseth} S.~J.,  1996, \mn@doi [Celestial Mechanics and
  Dynamical Astronomy] {10.1007/BF00728347}, \href
  {https://ui.adsabs.harvard.edu/abs/1996CeMDA..64..197M} {64, 197}

\bibitem[\protect\citeauthoryear{{Mikkola} \& {Merritt}}{{Mikkola} \&
  {Merritt}}{2006}]{2006MNRAS.372..219M}
{Mikkola} S.,  {Merritt} D.,  2006, \mn@doi [\mnras]
  {10.1111/j.1365-2966.2006.10854.x}, \href
  {http://cdsads.u-strasbg.fr/abs/2006MNRAS.372..219M} {372, 219}

\bibitem[\protect\citeauthoryear{{Mikkola} \& {Merritt}}{{Mikkola} \&
  {Merritt}}{2008}]{2008AJ....135.2398M}
{Mikkola} S.,  {Merritt} D.,  2008, \mn@doi [\aj]
  {10.1088/0004-6256/135/6/2398}, \href
  {http://adsabs.harvard.edu/abs/2008AJ....135.2398M} {135, 2398}

\bibitem[\protect\citeauthoryear{{Milankovitch}}{{Milankovitch}}{1939}]{milankovitch_39}
{Milankovitch} M.,  1939, Bull. Serb. Acad. Math. Nat., 6

\bibitem[\protect\citeauthoryear{{Moe} \& {Di Stefano}}{{Moe} \& {Di
  Stefano}}{2017}]{2017ApJS..230...15M}
{Moe} M.,  {Di Stefano} R.,  2017, \mn@doi [\apjs] {10.3847/1538-4365/aa6fb6},
  \href {http://adsabs.harvard.edu/abs/2017ApJS..230...15M} {230, 15}

\bibitem[\protect\citeauthoryear{{Musen}}{{Musen}}{1961}]{1961JGR....66.2797M}
{Musen} P.,  1961, \mn@doi [\jgr] {10.1029/JZ066i009p02797}, \href
  {http://cdsads.u-strasbg.fr/abs/1961JGR....66.2797M} {66, 2797}

\bibitem[\protect\citeauthoryear{{Naoz} \& {Fabrycky}}{{Naoz} \&
  {Fabrycky}}{2014}]{2014ApJ...793..137N}
{Naoz} S.,  {Fabrycky} D.~C.,  2014, \mn@doi [\apj]
  {10.1088/0004-637X/793/2/137}, \href
  {http://cdsads.u-strasbg.fr/abs/2014ApJ...793..137N} {793, 137}

\bibitem[\protect\citeauthoryear{{Naoz}, {Farr}  \& {Rasio}}{{Naoz}
  et~al.}{2012}]{2012ApJ...754L..36N}
{Naoz} S.,  {Farr} W.~M.,   {Rasio} F.~A.,  2012, \mn@doi [\apjl]
  {10.1088/2041-8205/754/2/L36}, \href
  {http://cdsads.u-strasbg.fr/abs/2012ApJ...754L..36N} {754, L36}

\bibitem[\protect\citeauthoryear{{Naoz}, {Farr}, {Lithwick}, {Rasio}  \&
  {Teyssandier}}{{Naoz} et~al.}{2013}]{2013MNRAS.431.2155N}
{Naoz} S.,  {Farr} W.~M.,  {Lithwick} Y.,  {Rasio} F.~A.,   {Teyssandier} J.,
  2013, \mn@doi [\mnras] {10.1093/mnras/stt302}, \href
  {http://adsabs.harvard.edu/abs/2013MNRAS.431.2155N} {431, 2155}

\bibitem[\protect\citeauthoryear{{Pejcha}, {Antognini}, {Shappee}  \&
  {Thompson}}{{Pejcha} et~al.}{2013}]{2013MNRAS.435..943P}
{Pejcha} O.,  {Antognini} J.~M.,  {Shappee} B.~J.,   {Thompson} T.~A.,  2013,
  \mn@doi [\mnras] {10.1093/mnras/stt1281}, \href
  {http://adsabs.harvard.edu/abs/2013MNRAS.435..943P} {435, 943}

\bibitem[\protect\citeauthoryear{{Pelupessy}, {van Elteren}, {de Vries},
  {McMillan}, {Drost}  \& {Portegies Zwart}}{{Pelupessy}
  et~al.}{2013}]{2013A&A...557A..84P}
{Pelupessy} F.~I.,  {van Elteren} A.,  {de Vries} N.,  {McMillan} S.~L.~W.,
  {Drost} N.,   {Portegies Zwart} S.~F.,  2013, \mn@doi [\aap]
  {10.1051/0004-6361/201321252}, \href
  {http://adsabs.harvard.edu/abs/2013A%26A...557A..84P} {557, A84}

\bibitem[\protect\citeauthoryear{{Perets} \& {Fabrycky}}{{Perets} \&
  {Fabrycky}}{2009}]{2009ApJ...697.1048P}
{Perets} H.~B.,  {Fabrycky} D.~C.,  2009, \mn@doi [\apj]
  {10.1088/0004-637X/697/2/1048}, \href
  {https://ui.adsabs.harvard.edu/abs/2009ApJ...697.1048P} {697, 1048}

\bibitem[\protect\citeauthoryear{{Perets} \& {Kratter}}{{Perets} \&
  {Kratter}}{2012}]{2012ApJ...760...99P}
{Perets} H.~B.,  {Kratter} K.~M.,  2012, \mn@doi [\apj]
  {10.1088/0004-637X/760/2/99}, \href
  {https://ui.adsabs.harvard.edu/abs/2012ApJ...760...99P} {760, 99}

\bibitem[\protect\citeauthoryear{{Petrovich}}{{Petrovich}}{2015}]{2015ApJ...799...27P}
{Petrovich} C.,  2015, \mn@doi [\apj] {10.1088/0004-637X/799/1/27}, \href
  {http://cdsads.u-strasbg.fr/abs/2015ApJ...799...27P} {799, 27}

\bibitem[\protect\citeauthoryear{{Petrovich} \& {Tremaine}}{{Petrovich} \&
  {Tremaine}}{2016}]{2016ApJ...829..132P}
{Petrovich} C.,  {Tremaine} S.,  2016, \mn@doi [\apj]
  {10.3847/0004-637X/829/2/132}, \href
  {http://adsabs.harvard.edu/abs/2016ApJ...829..132P} {829, 132}

\bibitem[\protect\citeauthoryear{{Portegies Zwart}, {McMillan}, {van Elteren},
  {Pelupessy}  \& {de Vries}}{{Portegies Zwart}
  et~al.}{2013}]{2013CoPhC.183..456P}
{Portegies Zwart} S.,  {McMillan} S.~L.~W.,  {van Elteren} E.,  {Pelupessy} I.,
    {de Vries} N.,  2013, \mn@doi [Computer Physics Communications]
  {10.1016/j.cpc.2012.09.024}, \href
  {http://adsabs.harvard.edu/abs/2013CoPhC.183..456P} {183, 456}

\bibitem[\protect\citeauthoryear{{Raghavan} et~al.,}{{Raghavan}
  et~al.}{2010}]{2010ApJS..190....1R}
{Raghavan} D.,  et~al., 2010, \mn@doi [\apjs] {10.1088/0067-0049/190/1/1},
  \href {http://adsabs.harvard.edu/abs/2010ApJS..190....1R} {190, 1}

\bibitem[\protect\citeauthoryear{{Randall} \& {Xianyu}}{{Randall} \&
  {Xianyu}}{2018a}]{2018ApJ...853...93R}
{Randall} L.,  {Xianyu} Z.-Z.,  2018a, \mn@doi [\apj]
  {10.3847/1538-4357/aaa1a2}, \href
  {https://ui.adsabs.harvard.edu/abs/2018ApJ...853...93R} {853, 93}

\bibitem[\protect\citeauthoryear{{Randall} \& {Xianyu}}{{Randall} \&
  {Xianyu}}{2018b}]{2018ApJ...864..134R}
{Randall} L.,  {Xianyu} Z.-Z.,  2018b, \mn@doi [\apj]
  {10.3847/1538-4357/aad7fe}, \href
  {https://ui.adsabs.harvard.edu/abs/2018ApJ...864..134R} {864, 134}

\bibitem[\protect\citeauthoryear{{Rantala}, {Pihajoki}, {Mannerkoski},
  {Johansson}  \& {Naab}}{{Rantala} et~al.}{2020}]{2020MNRAS.492.4131R}
{Rantala} A.,  {Pihajoki} P.,  {Mannerkoski} M.,  {Johansson} P.~H.,   {Naab}
  T.,  2020, \mn@doi [\mnras] {10.1093/mnras/staa084}, \href
  {https://ui.adsabs.harvard.edu/abs/2020MNRAS.492.4131R} {492, 4131}

\bibitem[\protect\citeauthoryear{{Rein} \& {Liu}}{{Rein} \&
  {Liu}}{2012}]{2012A&A...537A.128R}
{Rein} H.,  {Liu} S.~F.,  2012, \mn@doi [\aap] {10.1051/0004-6361/201118085},
  \href {https://ui.adsabs.harvard.edu/abs/2012A&A...537A.128R} {537, A128}

\bibitem[\protect\citeauthoryear{{Rein} \& {Spiegel}}{{Rein} \&
  {Spiegel}}{2015}]{2015MNRAS.446.1424R}
{Rein} H.,  {Spiegel} D.~S.,  2015, \mn@doi [\mnras] {10.1093/mnras/stu2164},
  \href {https://ui.adsabs.harvard.edu/abs/2015MNRAS.446.1424R} {446, 1424}

\bibitem[\protect\citeauthoryear{{Roa}}{{Roa}}{2017}]{2017rom..book.....R}
{Roa} J.,  2017, {Regularization in Orbital Mechanics; Theory and Practice}.
 Vol. 42

\bibitem[\protect\citeauthoryear{{Rosengren} \& {Scheeres}}{{Rosengren} \&
  {Scheeres}}{2014}]{2014CeMDA.118..197R}
{Rosengren} A.~J.,  {Scheeres} D.~J.,  2014, \mn@doi [Celestial Mechanics and
  Dynamical Astronomy] {10.1007/s10569-013-9530-7}, \href
  {http://cdsads.u-strasbg.fr/abs/2014CeMDA.118..197R} {118, 197}

\bibitem[\protect\citeauthoryear{{Schwamb} et~al.,}{{Schwamb}
  et~al.}{2013}]{2013ApJ...768..127S}
{Schwamb} M.~E.,  et~al., 2013, \mn@doi [\apj] {10.1088/0004-637X/768/2/127},
  \href {https://ui.adsabs.harvard.edu/abs/2013ApJ...768..127S} {768, 127}

\bibitem[\protect\citeauthoryear{{Silsbee} \& {Tremaine}}{{Silsbee} \&
  {Tremaine}}{2017}]{2017ApJ...836...39S}
{Silsbee} K.,  {Tremaine} S.,  2017, \mn@doi [\apj] {10.3847/1538-4357/aa5729},
  \href {http://adsabs.harvard.edu/abs/2017ApJ...836...39S} {836, 39}

\bibitem[\protect\citeauthoryear{{Stephan}, {Naoz}, {Ghez}, {Witzel},
  {Sitarski}, {Do}  \& {Kocsis}}{{Stephan} et~al.}{2016}]{2016MNRAS.460.3494S}
{Stephan} A.~P.,  {Naoz} S.,  {Ghez} A.~M.,  {Witzel} G.,  {Sitarski} B.~N.,
  {Do} T.,   {Kocsis} B.,  2016, \mn@doi [\mnras] {10.1093/mnras/stw1220},
  \href {http://adsabs.harvard.edu/abs/2016MNRAS.460.3494S} {460, 3494}

\bibitem[\protect\citeauthoryear{{Stiefel} \& {Scheifele}}{{Stiefel} \&
  {Scheifele}}{1971}]{1971lrcm.book.....S}
{Stiefel} E.~L.,  {Scheifele} G.,  1971, {Linear and regular celestial
  mechanics}.
Springer-Verlag, Berlin

\bibitem[\protect\citeauthoryear{{Thompson}}{{Thompson}}{2011}]{2011ApJ...741...82T}
{Thompson} T.~A.,  2011, \mn@doi [\apj] {10.1088/0004-637X/741/2/82}, \href
  {http://adsabs.harvard.edu/abs/2011ApJ...741...82T} {741, 82}

\bibitem[\protect\citeauthoryear{{Tokovinin}}{{Tokovinin}}{2014a}]{2014AJ....147...86T}
{Tokovinin} A.,  2014a, \mn@doi [\aj] {10.1088/0004-6256/147/4/86}, \href
  {http://adsabs.harvard.edu/abs/2014AJ....147...86T} {147, 86}

\bibitem[\protect\citeauthoryear{{Tokovinin}}{{Tokovinin}}{2014b}]{2014AJ....147...87T}
{Tokovinin} A.,  2014b, \mn@doi [\aj] {10.1088/0004-6256/147/4/87}, \href
  {http://adsabs.harvard.edu/abs/2014AJ....147...87T} {147, 87}

\bibitem[\protect\citeauthoryear{{Toonen}, {Perets}  \& {Hamers}}{{Toonen}
  et~al.}{2018}]{2018A&A...610A..22T}
{Toonen} S.,  {Perets} H.~B.,   {Hamers} A.~S.,  2018, \mn@doi [\aap]
  {10.1051/0004-6361/201731874}, \href
  {https://ui.adsabs.harvard.edu/abs/2018A&A...610A..22T} {610, A22}

\bibitem[\protect\citeauthoryear{{Tremaine}, {Touma}  \& {Namouni}}{{Tremaine}
  et~al.}{2009}]{2009AJ....137.3706T}
{Tremaine} S.,  {Touma} J.,   {Namouni} F.,  2009, \mn@doi [\aj]
  {10.1088/0004-6256/137/3/3706}, \href
  {http://cdsads.u-strasbg.fr/abs/2009AJ....137.3706T} {137, 3706}

\bibitem[\protect\citeauthoryear{{Vokrouhlick{\'y}}}{{Vokrouhlick{\'y}}}{2016}]{2016MNRAS.461.3964V}
{Vokrouhlick{\'y}} D.,  2016, \mn@doi [\mnras] {10.1093/mnras/stw1596}, \href
  {https://ui.adsabs.harvard.edu/abs/2016MNRAS.461.3964V} {461, 3964}

\bibitem[\protect\citeauthoryear{{Wang}, {Nitadori}  \& {Makino}}{{Wang}
  et~al.}{2020}]{2020MNRAS.tmp..452W}
{Wang} L.,  {Nitadori} K.,   {Makino} J.,  2020, \mn@doi [\mnras]
  {10.1093/mnras/staa480}, \href
  {https://ui.adsabs.harvard.edu/abs/2020MNRAS.tmp..452W} {}

\bibitem[\protect\citeauthoryear{{Winn} \& {Fabrycky}}{{Winn} \&
  {Fabrycky}}{2015}]{2015ARA&A..53..409W}
{Winn} J.~N.,  {Fabrycky} D.~C.,  2015, \mn@doi [\araa]
  {10.1146/annurev-astro-082214-122246}, \href
  {https://ui.adsabs.harvard.edu/abs/2015ARA&A..53..409W} {53, 409}

\bibitem[\protect\citeauthoryear{{Wu} \& {Murray}}{{Wu} \&
  {Murray}}{2003}]{2003ApJ...589..605W}
{Wu} Y.,  {Murray} N.,  2003, \mn@doi [\apj] {10.1086/374598}, \href
  {http://cdsads.u-strasbg.fr/abs/2003ApJ...589..605W} {589, 605}

\bibitem[\protect\citeauthoryear{{Zanazzi} \& {Lai}}{{Zanazzi} \&
  {Lai}}{2017}]{2017MNRAS.467.1957Z}
{Zanazzi} J.~J.,  {Lai} D.,  2017, \mn@doi [\mnras] {10.1093/mnras/stx208},
  \href {https://ui.adsabs.harvard.edu/abs/2017MNRAS.467.1957Z} {467, 1957}

\bibitem[\protect\citeauthoryear{{Zanazzi} \& {Lai}}{{Zanazzi} \&
  {Lai}}{2018}]{2018MNRAS.477.5207Z}
{Zanazzi} J.~J.,  {Lai} D.,  2018, \mn@doi [\mnras] {10.1093/mnras/sty951},
  \href {https://ui.adsabs.harvard.edu/abs/2018MNRAS.477.5207Z} {477, 5207}

\bibitem[\protect\citeauthoryear{{von Zeipel}}{{von
  Zeipel}}{1910}]{1910AN....183..345V}
{von Zeipel} H.,  1910, \mn@doi [Astronomische Nachrichten]
  {10.1002/asna.19091832202}, \href
  {https://ui.adsabs.harvard.edu/abs/1910AN....183..345V} {183, 345}

\makeatother
\end{thebibliography}

\appendix

\onecolumn

\section{Equations of motion from the octupole-order triplet Hamiltonian}
\label{app:H_triplet}
\subsection{Nonaveraged case}
\label{app:H_triplet:un}
The nonaveraged triplet Hamiltonian at octupole order, \eq~(\ref{eq:H_triplet}), gives a contribution to the equations of motion of the innermost orbit $p$ given by
\begin{align}
\nonumber &\ddot{\ve{r}}_{p}= - \frac{1}{\mu_p} \frac{ \partial S'_{3;3}(p,u,k)}{\partial \ve{r}_p} = - \frac{3}{2} \alpha(p,k.\mathrm{CS}(p);k) \, G M_{k.\mathrm{CS}(p)} \frac{\alpha(p,k.\mathrm{CS}(p);u) M_{u.\mathrm{CS}(p)}}{M_u} \Biggl [ 2\frac{\vdot{\ve{r}_p}{\ve{r}_k} \vdot{\ve{r}_u}{\ve{r}_k} \ve{r}_k}{r_k^7} - 2 \frac{\vdot{\ve{r}_p}{\ve{r}_u}\ve{r}_k + \vdot{\ve{r}_p}{\ve{r}_k}\ve{r}_u}{r_k^5} \\
&\quad - 2 \frac{\vdot{\ve{r}_u}{\ve{r}_k} \ve{r}_p}{r_k^5} \Biggl ].
\end{align}
The contribution to the intermediate orbit $u$ is
\begin{align}
\nonumber &\ddot{\ve{r}}_{u}= - \frac{1}{\mu_u} \frac{ \partial S'_{3;3}(p,u,k)}{\partial \ve{r}_u} = - \frac{3}{2} \frac{\mu_p}{\mu_u} \alpha(p,k.\mathrm{CS}(p);k) \,G M_{k.\mathrm{CS}(p)} \frac{\alpha(p,k.\mathrm{CS}(p);u) M_{u.\mathrm{CS}(p)}}{M_u} \Biggl [ \frac{\vdot{\ve{r}_p}{\ve{r}_k}^2 \ve{r}_k}{r_k^7} - 2 \frac{\vdot{\ve{r}_p}{\ve{r}_k} \ve{r}_p}{r_k^5} - \frac{r_p^2 \ve{r}_k}{r_k^5} \Biggl ].
\end{align}
Lastly, for the outermost orbit $k$,
\begin{align}
\nonumber &\ddot{\ve{r}}_{k}= - \frac{1}{\mu_k} \frac{ \partial S'_{3;3}(p,u,k)}{\partial \ve{r}_k} = - \frac{3}{2} \frac{\mu_p}{\mu_k} \alpha(p,k.\mathrm{CS}(p);k) \,G M_{k.\mathrm{CS}(p)} \frac{\alpha(p,k.\mathrm{CS}(p);u) M_{u.\mathrm{CS}(p)}}{M_u} \Biggl [\frac{2\vdot{\ve{r}_p}{\ve{r}_k} \vdot{\ve{r}_u}{\ve{r}_k}\ve{r}_p + \vdot{\ve{r}_p}{\ve{r}_k}^2 \ve{r}_u}{r_k^7} \\
\nonumber &\qquad - 7 \frac{\vdot{\ve{r}_p}{\ve{r}_k}^2 \vdot{\ve{r}_u}{\ve{r}_k}\ve{r}_k}{r_k^9} - 2 \frac{\vdot{\ve{r}_p}{\ve{r}_u} \ve{r}_p}{r_k^5} + 10 \frac{\vdot{\ve{r}_p}{\ve{r}_k}\vdot{\ve{r}_p}{\ve{r}_u} \ve{r}_k}{r_k^7} - \frac{r_p^2 \ve{r}_u}{r_k^5} + 5 \frac{\vdot{\ve{r}_u}{\ve{r}_k} r_p^2 \ve{r}_k}{r_k^7} \Biggl ].
\end{align}

\subsection{Inner-averaged case}
\label{app:H_triplet:A}
The inner-averaged triplet Hamiltonian, \eq~(\ref{eq:H_triplet_A}), gives a contribution to the equations of motion of the innermost orbit $p$ given by \eqs~(\ref{eq:EOM_Mil}), where the gradients are given by
\begin{subequations}
\begin{align}
\nonumber &\frac{\partial  \left \langle S'_{3;3}(p,u,k) \right \rangle_p}{\partial \ve{e}_p} = \frac{3}{4} \alpha(p,k.\mathrm{CS}(p);k) \,G M_{k.\mathrm{CS}(p)} \frac{\alpha(p,k.\mathrm{CS}(p);u) M_{u.\mathrm{CS}(p)}}{M_u} a_p^2 \Biggl [ \frac{\vdot{\ve{r}_u}{\ve{r}_k} \left \{ -2 r_k^2 \ve{e}_p + 10 \vdot{\ve{e}_p}{\ve{r}_k} \ve{r}_k \right \}}{r_k^7} \\
&\quad -2 \frac{-2\vdot{\ve{r}_u}{\ve{r}_k}\ve{e}_p + 5 \vdot{\ve{e}_p}{\ve{r}_k} \ve{r}_u + 5 \vdot{\ve{e}_p}{\ve{r}_u} \ve{r}_k}{r_k^5} -6 \frac{\vdot{\ve{r}_u}{\ve{r}_k}\ve{e}_p}{r_k^5} \Biggl ]; \\
&\frac{\partial  \left \langle S'_{3;3}(p,u,k) \right \rangle_p}{\partial \ve{\j}_p} = \frac{3}{4} \alpha(p,k.\mathrm{CS}(p);k) \,G M_{k.\mathrm{CS}(p)} \frac{\alpha(p,k.\mathrm{CS}(p);u) M_{u.\mathrm{CS}(p)}}{M_u} a_p^2 \Biggl [ -2 \frac{\vdot{\ve{r}_u}{\ve{r}_k} \vdot{\ve{\j}_p}{\ve{r}_k} \ve{r}_k}{r_k^7} + 2 \frac{\vdot{\ve{\j}_p}{\ve{r}_k} \ve{r}_u + \vdot{\ve{\j}_p}{\ve{r}_u}\ve{r}_k}{r_k^5} \Biggl ].
\end{align}
\end{subequations}
The (nonaveraged) intermediate orbit evolves according to
\begin{align}
\nonumber &\ddot{\ve{r}}_{u}= - \frac{1}{\mu_u} \frac{ \partial \left \langle S'_{3;3}(p,u,k) \right \rangle_p}{\partial \ve{r}_u} = - \frac{3}{4} \frac{\mu_p}{\mu_u} \alpha(p,k.\mathrm{CS}(p);k) \,G M_{k.\mathrm{CS}(p)} \frac{\alpha(p,k.\mathrm{CS}(p);u) M_{u.\mathrm{CS}(p)}}{M_u} a_p^2 \Biggl [ \frac{ \left(1-e_p^2\right) r_k^2 + 5 \vdot{\ve{e}_p}{\ve{r}_k}^2 - \vdot{\ve{\j}_p}{\ve{r}_k}^2}{r_k^7}\ve{r}_k \\
&\quad - 2 \frac{\left (1-e_p^2\right) \ve{r}_k + 5 \vdot{\ve{e}_p}{\ve{r}_k} \ve{e}_p - \vdot{\ve{\j}_p}{\ve{r}_k} \ve{\j}_p}{r_k^5} - \frac{\left(2+3e_p^2\right) \ve{r}_k}{r_k^5} \Biggl ],
\end{align}
and the outer orbit according to
\begin{align}
\nonumber &\ddot{\ve{r}}_{k}= - \frac{1}{\mu_k} \frac{ \partial \left \langle S'_{3;3}(p,u,k) \right \rangle_p}{\partial \ve{r}_k} = - \frac{3}{4} \frac{\mu_p}{\mu_k} \alpha(p,k.\mathrm{CS}(p);k) \,G M_{k.\mathrm{CS}(p)} \frac{\alpha(p,k.\mathrm{CS}(p);u) M_{u.\mathrm{CS}(p)}}{M_u} a_p^2 \\
\nonumber &\quad \times \Biggl [ \frac{ \ve{r}_u \left \{ \left(1-e_p^2\right) r_k^2 + 5 \vdot{\ve{e}_p}{\ve{r}_k}^2 - \vdot{\ve{\j}_p}{\ve{r}_k}^2 \right \} + \vdot{\ve{r}_u}{\ve{r}_k} \left \{ 2 \left(1-e_p^2\right)\ve{r}_k + 10 \vdot{\ve{e}_p}{\ve{r}_k} \ve{e}_p - 2 \vdot{\ve{\j}_p}{\ve{r}_k} \ve{\j}_p \right \}}{r_k^7} \\
\nonumber &\qquad - 7 \frac{\left(1-e_p^2\right) r_k^2 + 5 \vdot{\ve{e}_p}{\ve{r}_k}^2 - \vdot{\ve{\j}_p}{\ve{r}_k}^2}{r_k^9} \vdot{\ve{r}_u}{\ve{r}_k} \ve{r}_k - 2 \frac{\left(1-e_p^2\right) \ve{r}_u + 5 \vdot{\ve{e}_p}{\ve{r}_u} \ve{e}_p - \vdot{\ve{\j}_p}{\ve{r}_u} \ve{\j}_p}{r_k^5} \\
\nonumber &\qquad + 10 \frac{ \left(1-e_p^2\right) \vdot{\ve{r}_u}{\ve{r}_k} + 5 \vdot{\ve{e}_p}{\ve{r}_u} \vdot{\ve{e}_p}{\ve{r}_k} - \vdot{\ve{\j}_p}{\ve{r}_u} \vdot{\ve{\j}_p}{\ve{r}_k}}{r_k^7} \ve{r}_k - \frac{\left(2+3e_p^2\right) \ve{r}_u}{r_k^5} + 5 \frac{\vdot{\ve{r}_u}{\ve{r}_k} \left(2+3e_p^2\right) \ve{r}_k}{r_k^7} \Biggl ].
\end{align}

\subsection{Double-averaged case}
\label{app:H_triplet:AB}
The inner- and intermediate-averaged triplet Hamiltonian, \eq~(\ref{eq:H_triplet_AB}), gives a contribution to the equations of motion of the innermost orbit, $p$, given by \eq~(\ref{eq:EOM_Mil}), where the gradients are given by
\begin{subequations}
\begin{align}
\nonumber &\frac{\partial  \left \langle S'_{3;3}(p,u,k) \right \rangle_{p,u}}{\partial \ve{e}_p} = -\frac{9}{8} \alpha(p,k.\mathrm{CS}(p);k) \,G M_{k.\mathrm{CS}(p)} \frac{\alpha(p,k.\mathrm{CS}(p);u) M_{u.\mathrm{CS}(p)}}{M_u} a_p^2 a_u \Biggl [ \frac{\vdot{\ve{e}_u}{\ve{r}_k} \left \{ -2 r_k^2 \ve{e}_p + 10 \vdot{\ve{e}_p}{\ve{r}_k} \ve{r}_k \right \}}{r_k^7} \\
&\quad -2 \frac{-2\vdot{\ve{e}_u}{\ve{r}_k}\ve{e}_p + 5 \vdot{\ve{e}_p}{\ve{r}_k} \ve{e}_u + 5 \vdot{\ve{e}_p}{\ve{e}_u} \ve{r}_k}{r_k^5} -6 \frac{\vdot{\ve{e}_u}{\ve{r}_k}\ve{e}_p}{r_k^5} \Biggl ]; \\
&\frac{\partial  \left \langle S'_{3;3}(p,u,k) \right \rangle_{p,u}}{\partial \ve{\j}_p} = -\frac{9}{8} \alpha(p,k.\mathrm{CS}(p);k) \,G M_{k.\mathrm{CS}(p)} \frac{\alpha(p,k.\mathrm{CS}(p);u) M_{u.\mathrm{CS}(p)}}{M_u} a_p^2 a_u \Biggl [ -2 \frac{\vdot{\ve{e}_u}{\ve{r}_k} \vdot{\ve{\j}_p}{\ve{r}_k} \ve{r}_k}{r_k^7} + 2 \frac{\vdot{\ve{\j}_p}{\ve{r}_k} \ve{e}_u + \vdot{\ve{\j}_p}{\ve{e}_u}\ve{r}_k}{r_k^5} \Biggl ].
\end{align}
\end{subequations}
The averaged intermediate orbit evolves according to
\begin{subequations}
\begin{align}
\nonumber &\frac{\partial  \left \langle S'_{3;3}(p,u,k) \right \rangle_{p,u}}{\partial \ve{e}_u} = -\frac{9}{8} \alpha(p,k.\mathrm{CS}(p);k) \,G M_{k.\mathrm{CS}(p)} \frac{\alpha(p,k.\mathrm{CS}(p);u) M_{u.\mathrm{CS}(p)}}{M_u} a_p^2 a_u \Biggl [ \frac{\left(1-e_p^2\right) r_k^2 + 5 \vdot{\ve{e}_p}{\ve{r}_k}^2 - \vdot{\ve{\j}_p}{\ve{r}_k}^2}{r_k^7}\ve{r}_k \\
&\quad - 2 \frac{\left(1-e_p^2\right)\ve{r}_k + 5 \vdot{\ve{e}_p}{\ve{r}_k} \ve{e}_p - \vdot{\ve{\j}_p}{\ve{r}_k} \ve{\j}_p}{r_k^5} - \frac{\left(2+3e_p^2\right) \ve{r}_k}{r_k^5} \Biggl ]; \\
&\frac{\partial  \left \langle S'_{3;3}(p,u,k) \right \rangle_{p,u}}{\partial \ve{\j}_u} = \ve{0}.
\end{align}
\end{subequations}
Lastly, the nonaveraged outermost orbit evolves according to
\begin{align}
\nonumber &\ddot{\ve{r}}_{k}= - \frac{1}{\mu_k} \frac{ \partial S'_{3;3}(p,u,k)}{\partial \ve{r}_k} = \frac{9}{8} \frac{\mu_p}{\mu_k} \alpha(p,k.\mathrm{CS}(p);k) \,G M_{k.\mathrm{CS}(p)} \frac{\alpha(p,k.\mathrm{CS}(p);u) M_{u.\mathrm{CS}(p)}}{M_u} a_p^2 a_u \\
\nonumber &\quad \times \Biggl [ \frac{ \ve{r}_u \left \{ \left(1-e_p^2\right) r_k^2 + 5 \vdot{\ve{e}_p}{\ve{r}_k}^2 - \vdot{\ve{\j}_p}{\ve{r}_k}^2 \right \} + \vdot{\ve{e}_u}{\ve{r}_k} \left \{ 2 \left(1-e_p^2\right)\ve{r}_k + 10 \vdot{\ve{e}_p}{\ve{r}_k} \ve{e}_p - 2 \vdot{\ve{\j}_p}{\ve{r}_k} \ve{\j}_p \right \}}{r_k^7} \\
\nonumber &\qquad - 7 \frac{\left(1-e_p^2\right) r_k^2 + 5 \vdot{\ve{e}_p}{\ve{r}_k}^2 - \vdot{\ve{\j}_p}{\ve{r}_k}^2}{r_k^9} \vdot{\ve{e}_u}{\ve{r}_k} \ve{r}_k - 2 \frac{\left(1-e_p^2\right) \ve{e}_u + 5 \vdot{\ve{e}_p}{\ve{e}_u} \ve{e}_p - \vdot{\ve{\j}_p}{\ve{e}_u} \ve{\j}_p}{r_k^5} \\
\nonumber &\qquad + 10 \frac{ \left(1-e_p^2\right) \vdot{\ve{e}_u}{\ve{r}_k} + 5 \vdot{\ve{e}_p}{\ve{e}_u} \vdot{\ve{e}_p}{\ve{r}_k} - \vdot{\ve{\j}_p}{\ve{e}_u} \vdot{\ve{\j}_p}{\ve{r}_k}}{r_k^7} \ve{r}_k - \frac{\left(2+3e_p^2\right) \ve{e}_u}{r_k^5} + 5 \frac{\vdot{\ve{e}_u}{\ve{r}_k} \left(2+3e_p^2\right) \ve{r}_k}{r_k^7} \Biggl ].
\end{align}

\label{lastpage}

\end{document}